\documentclass[11pt,a4paper]{article}
\usepackage{times,latexsym}
\usepackage{url}
\usepackage[table,xcdraw]{xcolor}
\usepackage{pifont} 
\usepackage{listings}
\usepackage{graphicx}
\usepackage{booktabs}
\usepackage{amsmath}
\usepackage{subcaption}
\usepackage[most]{tcolorbox}
\usepackage{lipsum} % solo per testo di esempio
\usepackage{fontawesome5}
\usepackage[T1]{fontenc}

\usepackage[acceptedWithA]{tacl2021v1}

\usepackage{xspace,mfirstuc,tabulary}

\newif\iftaclinstructions
\taclinstructionstrue
\iftaclinstructions
\newcommand{\instr}
\fi

\iftaclpubformat

\else

\fi

\title{AMAQA: A Metadata-based QA Dataset for RAG Systems}

\author{
  Davide Bruni$^\diamond$$^\ddagger$,
  Marco Avvenuti$^\dagger$,
  Nicola Tonellotto$^\dagger$,
  Maurizio Tesconi$^\diamond$\\
  $^\diamond$
Institute for Informatics and Telematics, National Research Council, Italy\\
\texttt{davide.bruni@iit.cnr.it, 	maurizio.tesconi@iit.cnr.it}\\
  $^\dagger$Department of Information Engineering, University of Pisa, Italy\\
  \texttt{marco.avvenuti@unipi.it, nicola.tonellotto@unipi.it}\\
  $^\ddagger$Department of Computer Science, University of Pisa, Italy\\
  \texttt{davide.bruni@phd.unipi.it}
}

\begin{document}
\maketitle
\begin{abstract}
Retrieval-augmented generation (RAG) systems are widely used in question-answering (QA) tasks, but current benchmarks lack metadata integration, limiting their evaluation in scenarios requiring both textual data and external information. To address this, we present AMAQA, a new open-access QA dataset designed to evaluate tasks combining text and metadata\footnote{The benchmark is available at \url{https://github.com/DavideBruni/AMAQA}}. The integration of metadata is especially important in fields that require rapid analysis of large volumes of data, such as cybersecurity and intelligence, where timely access to relevant information is critical. AMAQA includes about 1.1 million English messages collected from 26 public Telegram groups, enriched with metadata such as timestamps and chat names. It also contains 20,000 hotel reviews with metadata. In addition, the dataset provides 2,600 high-quality QA pairs built across both domains, Telegram messages and hotel reviews, making AMAQA a valuable resource for advancing research on metadata-driven QA and RAG systems. Both Telegram messages and Hotel reviews are enriched with emotional tones or  toxicity indicators. To the best of our knowledge, AMAQA is the first single-hop QA benchmark to incorporate metadata. We conduct extensive tests on the benchmark, setting a new reference point  for future research. We show that leveraging metadata boosts accuracy from 0.5 to 0.86 for GPT-4o and from 0.27 to 0.76 for open source LLMs, highlighting the value of structured context. We conducted experiments on our benchmark to assess the performance of known techniques designed to enhance RAG, highlighting the importance of properly managing metadata throughout the entire RAG pipeline.
\end{abstract}

\section{Introduction}
\begin{figure}[ht]
  %\centering
  \includegraphics[width=0.5\textwidth]{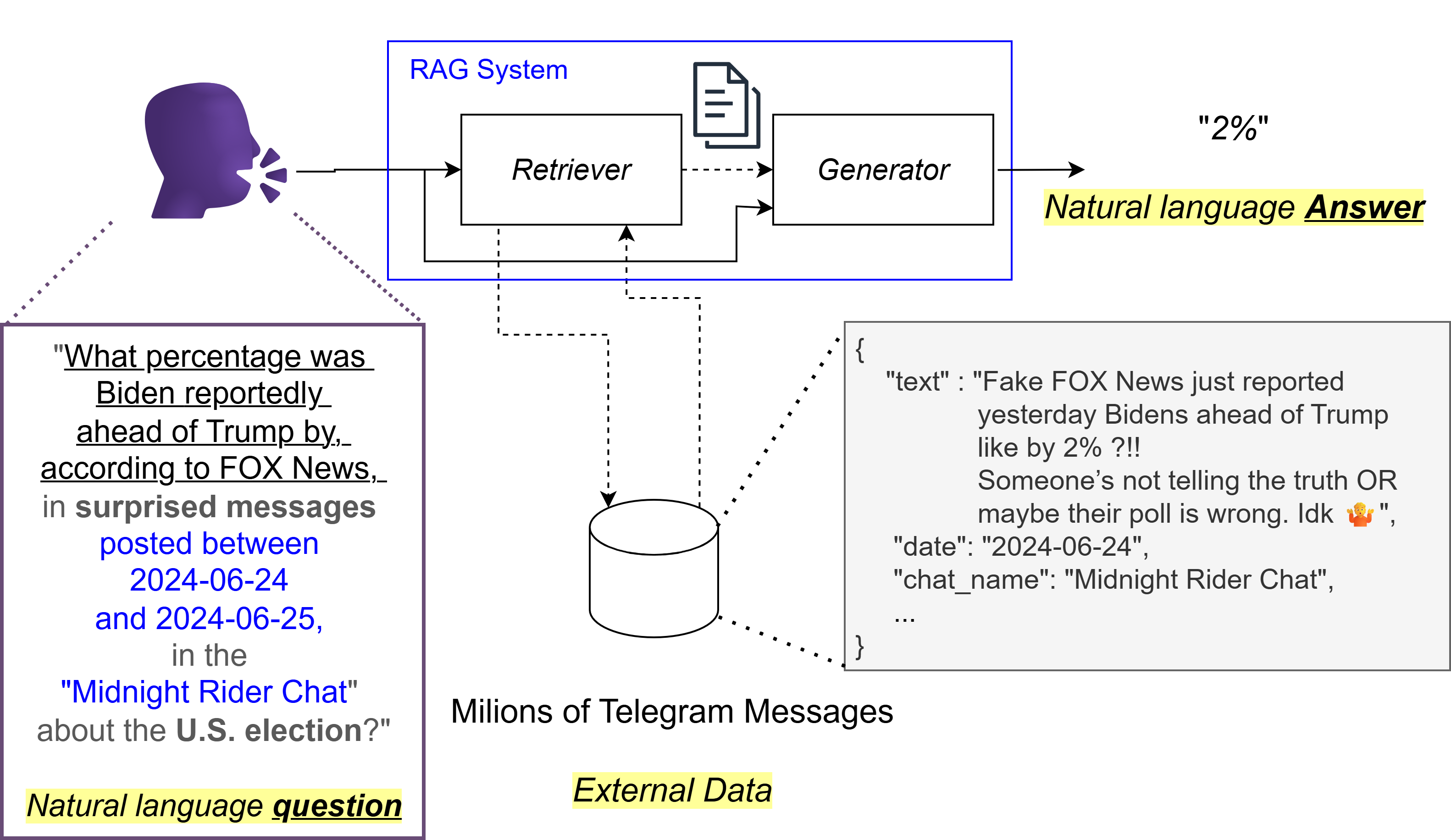}
  \caption{High-level conceptual schema of the systems analyzed in this work}
  \label{fig:metodology_schema}
\end{figure}

Retrieval-augmented generation (RAG) systems are increasingly utilized in knowledge-driven applications, including question answering, information extraction, and dialogue generation~\cite{gao2023retrieval}. While existing benchmarks have significantly advanced the evaluation of RAG systems, they primarily focus on text-based inputs, overlooking the potential of metadata, such as engagement metrics or message timestamps, alongside labels like topics, emotional tone, or toxicity.
However, the ability to use metadata effectively is particularly important in fields where the context surrounding textual data is as important as the content itself~\cite{pelofske2023cybersecurity,perez2018you}. For example, analysts often must identify relevant information from huge datasets by filtering contents based on metadata. For instance analysts could identify relevant information by filtering messages based on the time period or channel where the message was posted, and the tone or intent of the text itself. Consider a scenario similar to Figure~\ref{fig:metodology_schema}, in which an analyst must answer a question such as:
\begin{quote}
“What percentage was Biden reportedly ahead of Trump by, according to FOX News, in surprised messages posted between 2024-06-24 and 2024-06-25, in the “Midnight Rider Chat" about the U.S. election?"
\end{quote}

This question involves several layers of reasoning:
\begin{itemize}
\item \textit{Information extraction:}
The system must detect quantitative statements about polling data and extract the percentage by which Biden was reportedly ahead, while also recognizing FOX News as the cited source.
\item \textit{Metadata filters:}
The phrases \textit{“posted between 2024-06-24 and 2024-06-25"} and \textit{“in the Midnight Rider Chat"} translate directly into metadata constraints narrowing down the set of relevant messages.
\item \textit{Textual features not based on metadata:} Two other components of the question cannot be resolved solely using metadata: \textit{“in surprised messages"} requires the system to automatically recognize messages that express surprise. While \textit{“talking about U.S. election"} requires the system to detect messages related to the U.S. election, even without explicit mentions. Although emotional, toxicity and topic labels are present in the dataset, they are used only for evaluation, not for filtering.
\end{itemize}
The system must process textual features alongside the metadata constraints to identify the relevant subset of messages. The system’s response to the question is \textit{“2\%"}, which is correct because the dataset includes a June 24, 2024 message from the “Midnight Rider Chat” expressing surprise about the U.S. election and stating that, “according to FOX News, Biden was reportedly ahead of Trump by 2\%.". Although this may represent misinformation, it's correct since it accurately reflects the dataset. The system’s task is to infer this answer from available data, not to fact-check its accuracy. This structure emphasizes the need for a system capable of combining structured metadata processing with semantic text understanding, even when dealing with implicitly expressed content.

In this paper, we introduce AMAQA, a novel dataset designed to address the aforementioned gaps, and we evaluate the performance of four architectures (Vanilla RAG, RAG with metadata filtering, Vanilla RAG + reranker Re²G~\cite{glass-etal-2022-re2g}, and RAG with metadata filtering with reranker, which we refer to as Re²G* for brevity) on the new benchmark. Furthermore, we demonstrate the importance of leveraging metadata to enhance the performance of RAG systems when such information is available, and we highlight the remaining challenges in developing robust metadata-driven question-answering models. We provide a baseline for future research, enabling the development of more advanced and effective metadata-driven question-answering systems.

\section{Related work}
The landscape of research datasets and benchmarks for QA systems and RAG models has evolved significantly in recent years. Existing work can be categorized into two broad classes: (i) datasets containing textual data enriched with metadata and (ii) benchmark datasets tailored to specific aspects of QA or RAG systems but lacking in metadata integration. As shown in Table \ref{tab:datasets}, a wide variety of datasets contain textual data with associated metadata exist, such as \textit{Movie-Dataset}~\cite{banik_movies_dataset}, \textit{Hotel-Reviews}~\cite{datafiniti_hotel_reviews_2016}, \textit{Amazon-Reviews}~\cite{hou2024bridging}, \textit{Pushshift}~\cite{baumgartner2020pushshift}, and \textit{TgDataset}~\cite{10.1145/3690624.3709397}. These datasets provide rich textual content and corresponding metadata, enabling tasks such as sentiment analysis, recommendation systems, or exploring the Telegram ecosystem. However, they aren't suited for QA benchmarking, lacking structured question-answer pairs. Furthermore, these datasets do not contain information on the discussed topics, the emotions conveyed, or the toxicity of the text, missing key elements that could enhance the depth of analysis in QA models.

Conversely, several QA benchmarks, such as \textit{PopQA} ~\cite{mallen-etal-2023-trust}, \textit{TriviaQA}~\cite{joshi-etal-2017-triviaqa} and \textit{SQuAD 1.1}~\cite{rajpurkar-etal-2016-squad}, have been introduced to address specific challenges QA tasks. For example, \textit{HotpotQA}~\cite{yang-etal-2018-hotpotqa} evaluates a system's ability to retrieve and reason over multiple documents, emphasizing \textit{multi-hop reasoning}. \textit{GPQA}~\cite{rein2024gpqa}, on the other hand, provides a multiple-choice QA dataset with expert-validated questions in biology, physics, and chemistry. While valuable for evaluating a system’s domain-specific capabilities, it lacks metadata integration and contextual information. Another prominent benchmark, \textit{CoQA}~\cite{reddy2019coqa}, is designed for conversational QA, evaluating systems' ability to maintain context and coherence across multiple dialogue turns. However, like other QA benchmarks designed to evaluate RAG-addressable tasks, such as \textit{CRUD}~\cite{lyu2024crud}, \textit{CoQA}~\cite{reddy2019coqa} does not leverage metadata to provide additional context or structure to question answering.

The main limitation we aim to highlight in Table~\ref{tab:datasets} is that although these datasets are either large-scale or domain-specific, none of them include (nor exploit) metadata. Whether they contain millions of QA pairs, such as \textit{MS-MARCO}~\cite{nguyen2016ms} and \textit{MIRAGE}~\cite{xiong-etal-2024-benchmarking}, or focus on specific domains like \textit{GPQA}~\cite{rein2024gpqa}, they lack structured metadata that could provide deeper insights and improve model evaluation. To address these gaps, \textit{AMAQA} aims to serve as an innovative resource for QA and RAG systems. By integrating metadata with a structured QA dataset, \textit{AMAQA} will enable more holistic evaluations, considering the interplay between text and metadata. This approach will open new possibilities for \textit{metadata-filtering}, bridging the gap between general textual datasets and metadata-free QA benchmarks.

\begin{table}[ht]
    \resizebox{0.48\textwidth}{!}{%
    \begin{tabular}{llccc}
        \toprule
        Type & Name & QA & Docs & Metadata \\
        \midrule
        Bench & CRUD & \textasciitilde 3.2k & \textasciitilde 80k & \ding{55} \\
        Bench & PopQA & \textasciitilde14k & N/A & \ding{55}\\
        Bench & HotpotQA & \textasciitilde113k & N/A & \ding{55}\\
        Bench & TriviaQA & \textasciitilde 650k & \textasciitilde 662k & \ding{55}\\
        Bench & SQuAD 1.1 & \textasciitilde 100k & \textasciitilde 500 & \ding{55}\\
        Bench & MS-MARCO & \textasciitilde 1.1M & \textasciitilde 138M & \ding{55}\\
        Bench & MIRAGE & \textasciitilde 7.6k & \textasciitilde 60M & \ding{55}\\
        Bench & GPQA & 448 & N/A & \ding{55}\\
        Bench & CoQA & \textasciitilde 127k & \textasciitilde 7k & \ding{55}\\
        Data & Movie-Dataset & \ding{55} & 45k & \ding{51}\\
        Data & Hotel-Rev. & \ding{55} & 20k & \ding{51}\\
        Data & Amazon-Rev. & \ding{55} & \textasciitilde 571M & \ding{51}\\
        Data & Pushshift & \ding{55} & \textasciitilde 317M & \ding{51}\\
        Data & TgDataset & \ding{55} & \textasciitilde 400M & \ding{51}\\
        \hline
        \rowcolor{gray!20} Bench & AMAQA & 2600 & \textasciitilde 1.2M & \ding{51}\\
        \bottomrule
    \end{tabular}%
    }
    \caption{Overview of single-hop QA benchmarks (Bench) and textual datasets (Data).}
    \label{tab:datasets}
\end{table}

\section{The AMAQA dataset}
The AMAQA dataset consists of approximately 1.1 million Telegram messages with their associated metadata and 20 thousand hotel reviews taken from the Hotel reviews dataset~\cite{datafiniti_hotel_reviews_2016}. Since not all of the metadata provided by Telegram and by Hotel Reviews dataset are useful for creating our benchmark, only a subset was used. In addition, each Telegram message was labeled with the topics discussed, the main emotion conveyed, and the probabilities of containing toxicity, profanity, insults, identity attacks, or threats, while the Hotel reviews were labeled only with the main emotion conveyed.

\subsection{Messages collection}
To build a large English-language text dataset with metadata, messages were collected from Telegram because it is a popular messaging app with over 900 million users as of 2024~\cite{businessofapps_telegram_2025} and it offers easy access to public channel messages and related discussion groups. The data collection started from well-known channels discussing the following topics: \textit{the Russian-Ukrainian conflict, the U.S. election, and the Israeli-Palestinian conflict}. Then, using Telegram’s recommendation system, channels were selected based on relevance, English-language content, and availability of discussion groups. In total, 26 channels and their groups were identified. Messages were collected between June 13 and August 13, 2024, resulting in approximately \textit{1.1 million messages}.

\subsection{Data labeling}
The first step required to identify which were the characteristics of the texts to focus on: the first one were the \textit{emotions} conveyed by the text. We adopted Ekman’s model of six basic emotions~\cite{ekman1992argument} which is widely used in emotion detection research, where emotions are typically categorized in a multi-class classification framework~\cite{seyeditabari2018emotion}. The Ekman's model highlights sadness, happiness, anger, fear, disgust, and surprise as universally recognized emotions. To this set, we added a “neutral" category to account emotionally-free texts. Emotions were extracted using a Zero-Shot Classifier ~\cite{laurer_less_2023,plaza-del-arco-etal-2022-natural}, applying the hypothesis \textit{“This text expresses a feeling of..."} with the seven emotions as classes. In addition to emotional features, other textual characteristics, such as Toxicity, were extracted for Telegram messages. Toxicity is defined by the Perspective API~\cite{perspective_api} as “a rude, disrespectful, or unreasonable comment that is likely to make you leave a discussion". However, the use of the Perspective API was not limited to detecting toxicity alone but also extended to other forms of harmful language, including Identity Attack, Insult, Profanity, and Threat. Following the guidelines, we consider a comment to exhibit these forms of harmful language if the corresponding score exceeds a threshold of 0.7, as adopted by~\citet{alvisi2025mappingitaliantelegramecosystem}. It is worth noting that these additional features were extracted only for Telegram messages. Although some hotel reviews contained negative emotions, such as anger, they rarely included toxic language. Therefore, adding these features to the hotel reviews would have been largely uninformative, as the corresponding values would have been consistently empty. In addition to analyzing the emotions and sentiments expressed in Telegram messages, it was essential to extract the discussed \textit{topics} in order to build a comprehensive ground truth. The topic detection process was structured with the goal of associating multiple topics to each message ~\cite{wang-etal-2023-text2topic} taking care not to identify too many or too few topics. Initially, the dataset was first manually explored to identify an initial set of topics. BERTTopic was then used to improve topic identification~\cite{grootendorst2022bertopic}. However, since BERTTopic assigns only one topic per post and often generates clusters that are either too general or too specific, a hybrid approach was adopted. The manually defined topics were compared with BERTTopic results, and frequently occurring topics (over 1,000 posts) not already identified were added. The final topic set includes 58 topics covering a wide range of subjects in the dataset. Since AMAQA contains almost 1.2 million Telegram messages, manual labeling was unfeasible. Instead, we used GPT-4o~\cite{openai2024gpt4ocard} for topic extraction, as it has outperformed crowd workers in annotation tasks~\cite{gilardi2023chatgpt}. Each message was processed by GPT-4o using a prompt inspired by prior topic extraction approaches~\cite{mu-etal-2024-large, munker2024zero}, allowing from 0 to $N$ topics extracted per message. The output was a list of relevant topics per message. However, since LLMs do not always follow instructions with complete precision ~\cite{zhou2023instruction}, we cleaned the detected topic to correct inconsistencies (e.g., “states/Europe" to “organizations/European Union") and exclude valid but off-list topics (e.g., “states/Italy”) due to their low frequency.\\
The step of adding the discussed topic to the ground-truth was unnecessary for hotel reviews, as the nature of these texts made such an analysis largely redundant, offering little new information beyond what was already apparent, while adding unnecessary complexity to the dataset.

\subsection{Telegram messages statistics}
\begin{figure*}
    \begin{subfigure}[b]{0.48\textwidth}
        \includegraphics[width=1\textwidth]{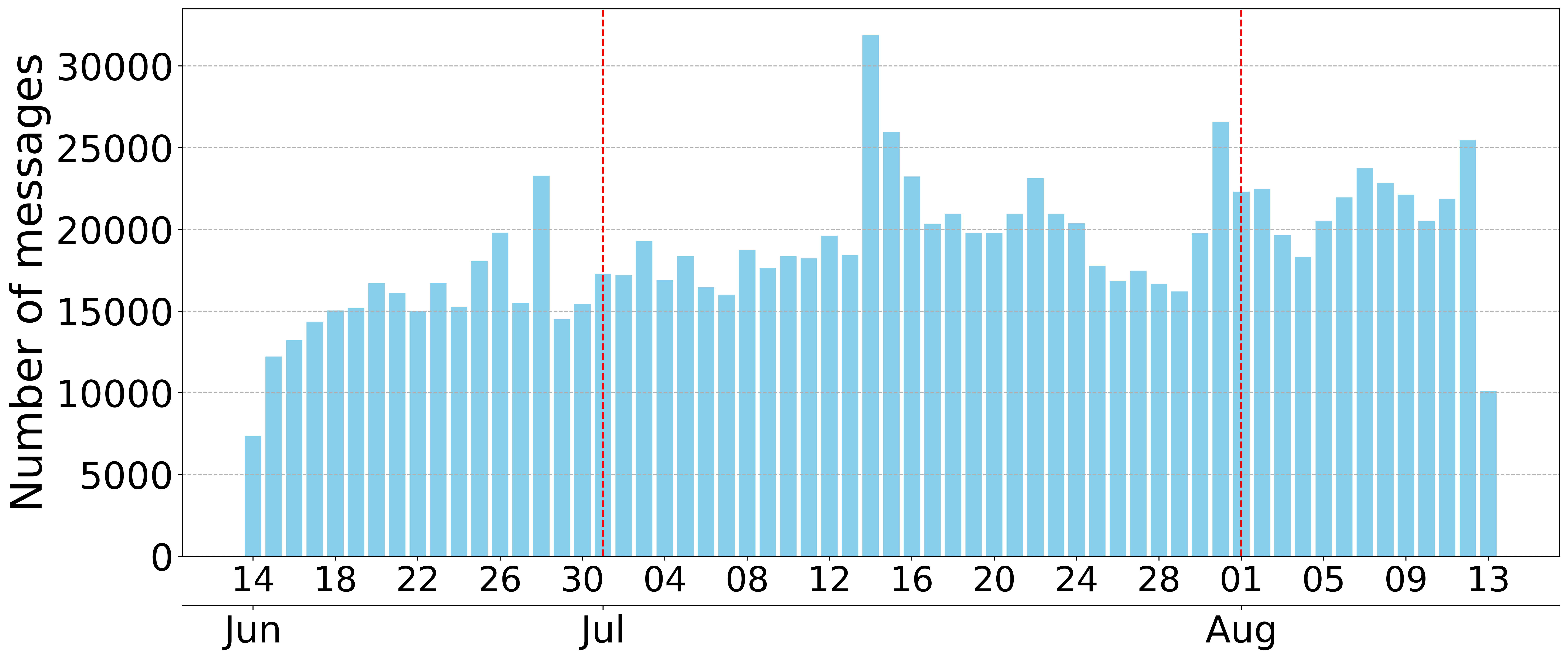}
        \caption{Distribution of messages over time}
        \label{fig:distributions_a}
    \end{subfigure}
    \begin{subfigure}[b]{0.48\textwidth}
        \includegraphics[width=1\textwidth]{topic_distribution.png}
        \caption{Distribution of the most frequently discussed topics}
        \label{fig:distributions_b}
    \end{subfigure}
    \caption{Visualization of message and topic distributions, providing insights into temporal trends and the division of messages by topic.}
\end{figure*}

\begin{figure}
  \centering
  \includegraphics[width=0.35\textwidth]{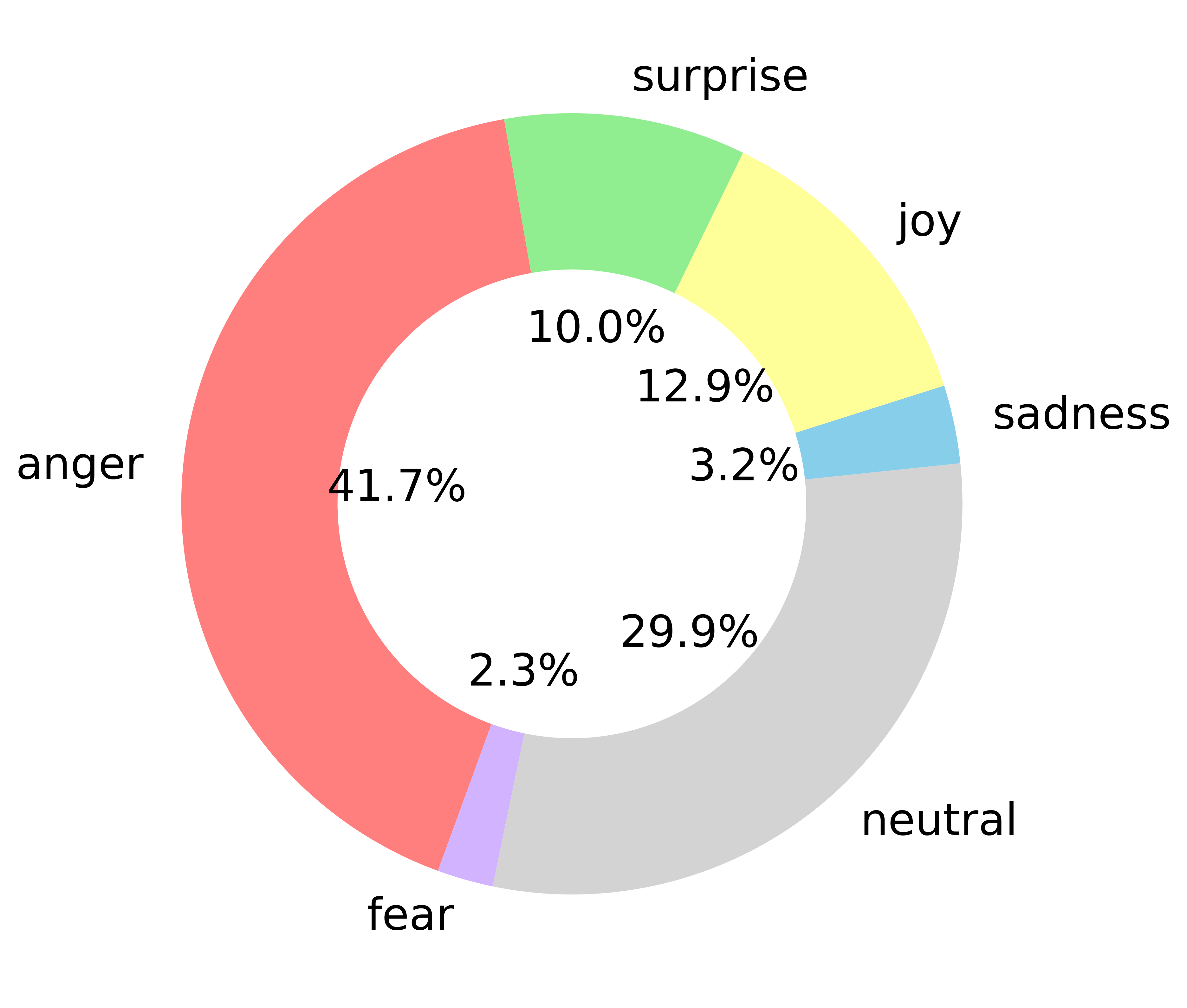}
  \caption{Distribution of emotions across messages, illustrating the prevalence of different emotional tones.}
  \label{fig:distributions_c}
\end{figure}

Table~\ref{tab:dataset_characteristics} summarizes the main characteristics of the Telegram messages in the dataset.
\begin{table}
    \centering
    \begin{tabular}{lr}
        \toprule
        \textit{Feature}             & \textit{Value} \\
        \midrule
        Total Messages               & 1,146,690      \\
        Toxic Messages               & 107,154        \\
        Contains Insults             & 52,413         \\
        Profanity                    & 61,969         \\
        Identity Attacks             & 9,239          \\
        Threats                      & 2,509          \\
        Groups                       & 26             \\
        Topics                       & 58             \\
        \bottomrule
    \end{tabular}
    \caption{Telegram messages characteristics}
    \label{tab:dataset_characteristics}
\end{table}
Additional insights are provided through visualizations: in Figure ~\ref{fig:distributions_a} the distribution of messages collected is shown. Notably, the temporal distribution shows activity peaks linked to major events, such as the July 14 attack on Trump and the June 28 Trump–Biden debate, while in Figure ~\ref{fig:distributions_c} a donut chart illustrates the distribution of emotions.

Figure~\ref{fig:distributions_b} highlights the main topics discussed in the dataset, with \textit{Russia}, \textit{Donald Trump}, \textit{Ukraine} and the \textit{Ukraine-Russia conflict} being the most prominent. This distribution is expected, as the seed channels primarily focus on these issues. The dataset is clearly unbalanced in terms of topic, emphasizing geopolitical themes and political figures such as states, conflicts, and leaders like Trump, Biden, and Putin. Other recurring topics include religion, media manipulation, and references to socially and politically sensitive groups, such as Muslims, Jews, and Democrats, all contributing to a highly polarized discourse. This is reflected in the emotional analysis, which shows a predominance of anger, suggesting intense reactions to political and conflict-related topics. Although joy is the second most common emotion, it appears far less frequently, reinforcing the dataset’s overall negative emotional tone. The prevalence of anger and focus on divisive themes confirm the polarized nature of the dataset, consistent with our expectations. The dataset was deliberately curated to include toxic content, such as threats and identity attacks, to facilitate the exploration of these textual characteristics. The interplay of complex themes and emotionally charged discussions underscores a landscape of intense and divisive debates.

\subsection{Hotel reviews statistics}
\begin{figure}
  \centering
  \includegraphics[width=0.48\textwidth]{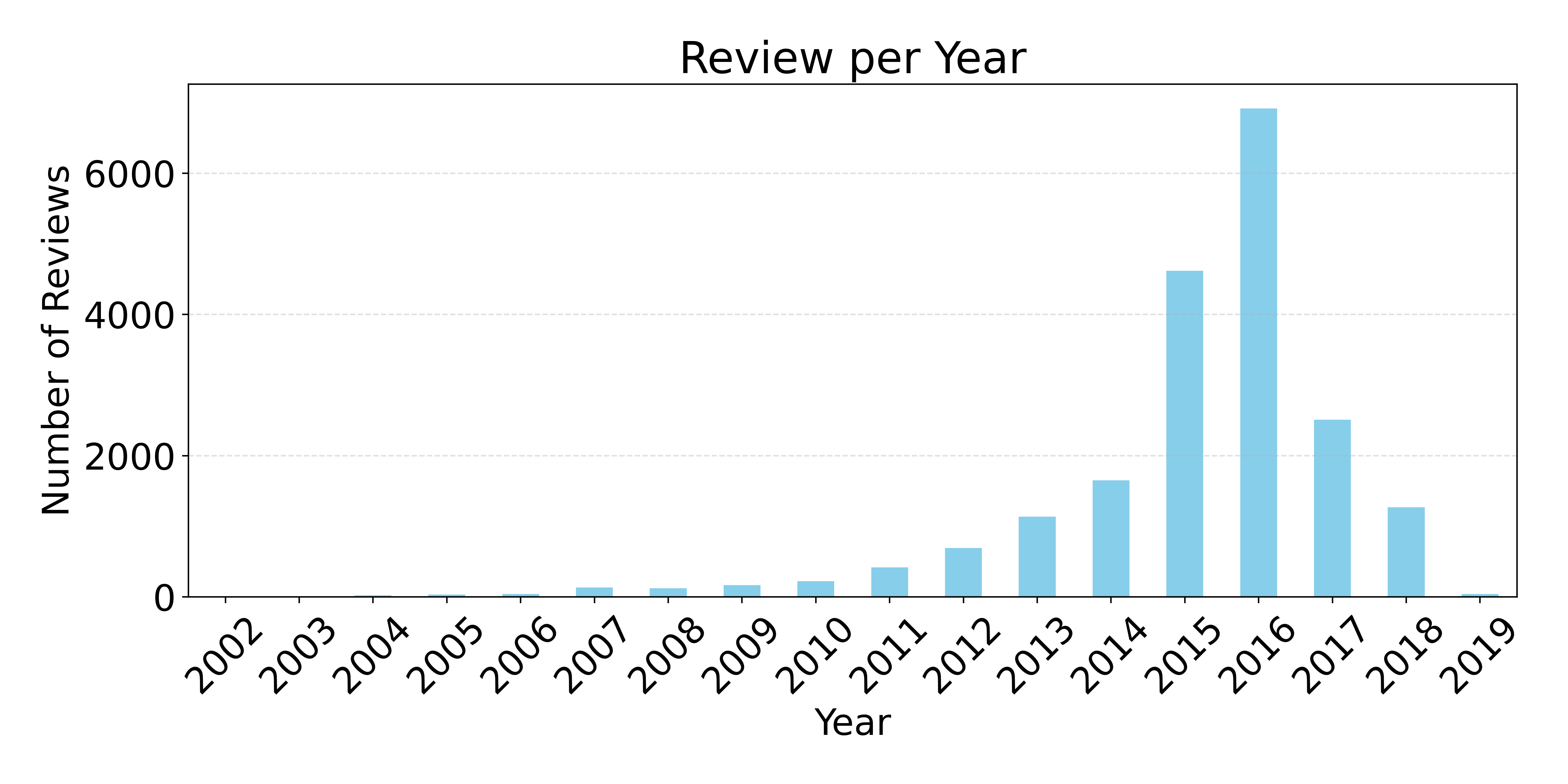}
  \caption{Distribution of hotel reviews over years.}
  \label{fig:hr_distribution}
\end{figure}

\begin{figure}[ht]
  \centering
  \includegraphics[width=0.35\textwidth]{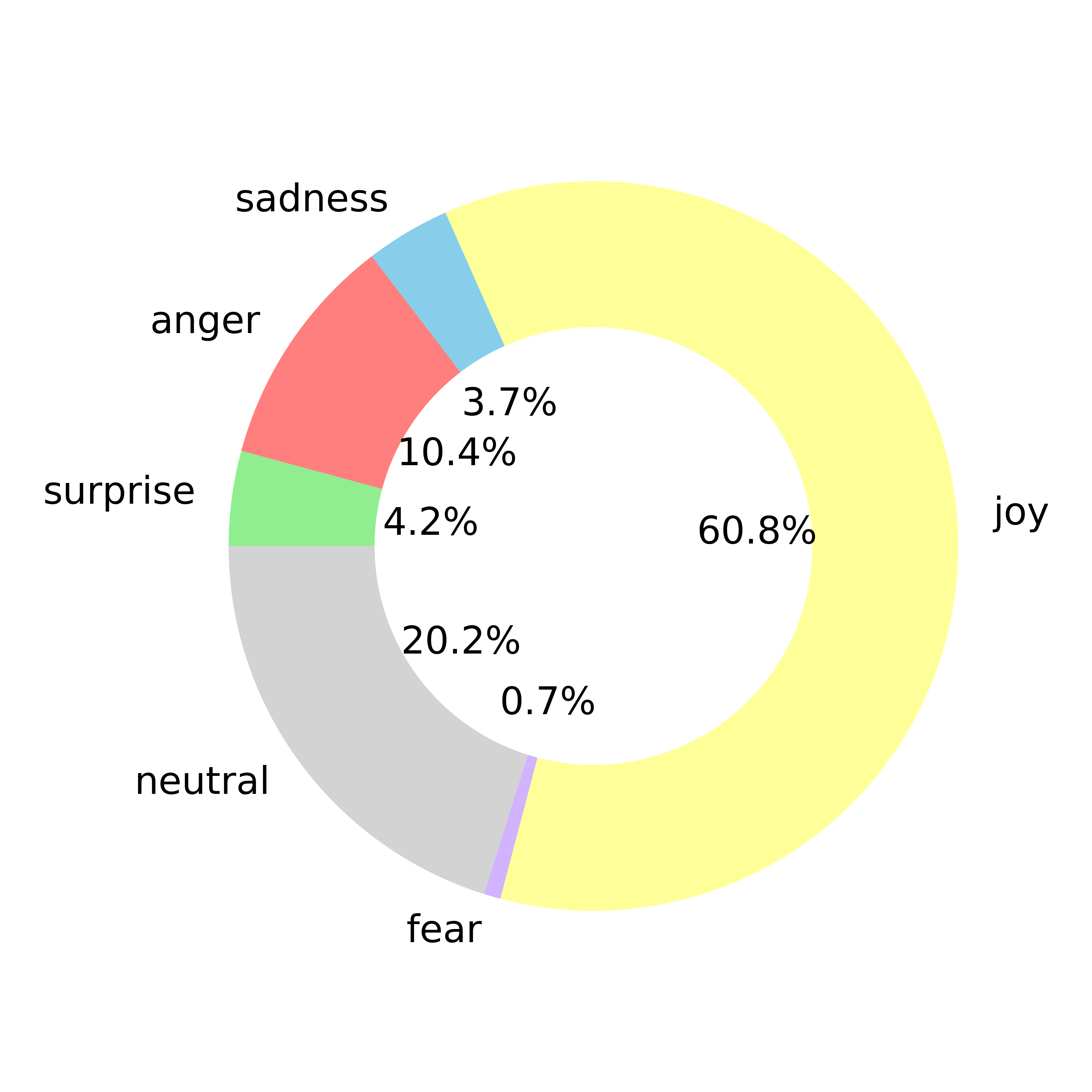}
  \caption{Distribution of emotions across hotel reviews, illustrating the prevalence of different emotional tones.}
  \label{fig:emotions_hotel}
\end{figure}

\begin{figure}[ht]
  \centering
  \includegraphics[width=0.48\textwidth]{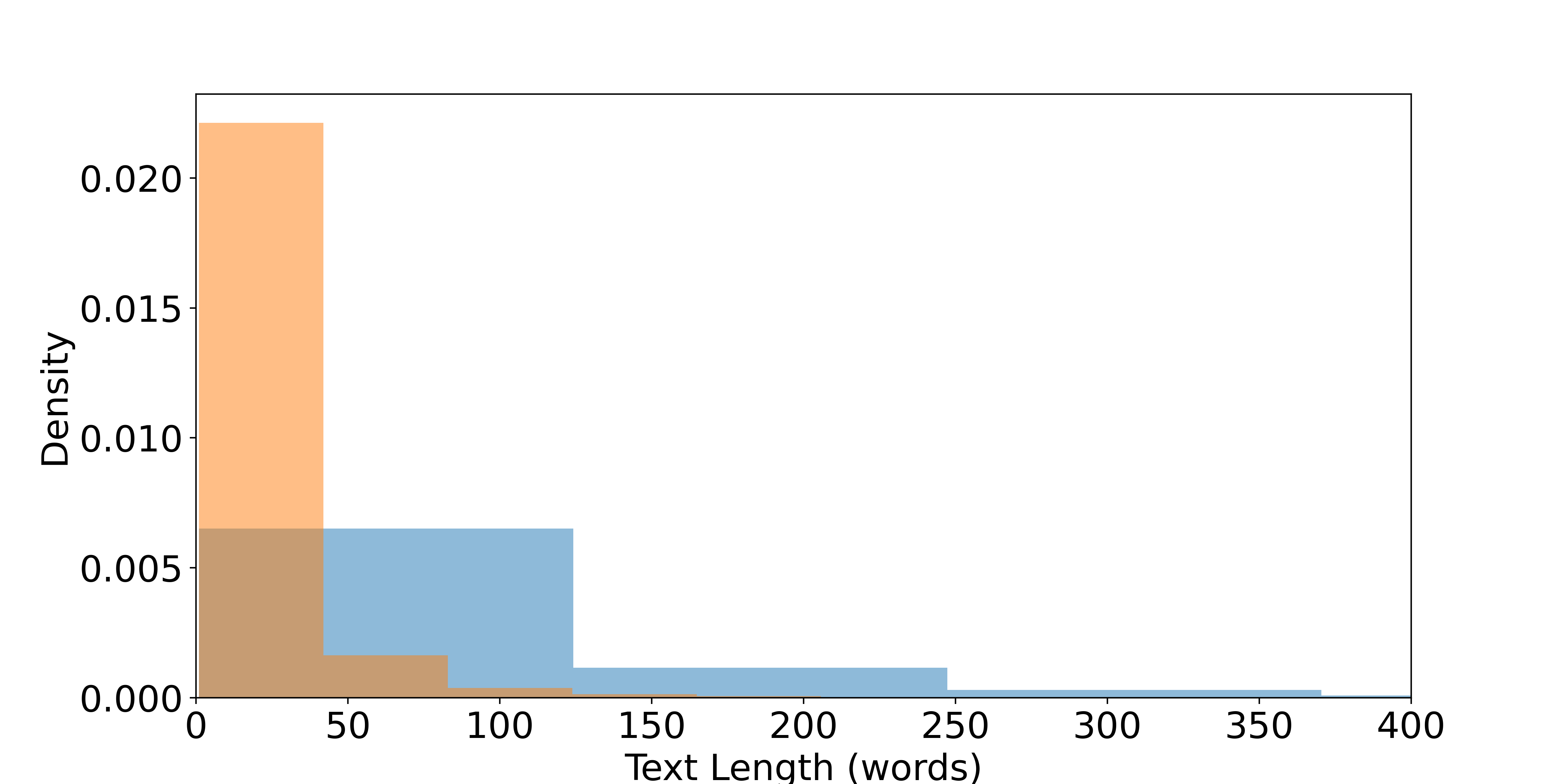}
  \caption{Normalized distribution of text lengths for Hotel Reviews and Telegram messages.}
  \label{fig:text_length}
\end{figure}

Table~\ref{tab:hr_characteristics} summarizes the key characteristics of the hotel reviews subset.
\begin{table}[ht]
    \centering
    \begin{tabular}{lr}
        \toprule
        \textit{Feature}             & \textit{Value} \\
        \midrule
        Total Reviews                & 20,000         \\
        Hotels                       & 2,764         \\
        Cities                       & 1,416          \\
        Provinces                    & 50             \\
        \bottomrule
    \end{tabular}
    \caption{Hotel reviews characteristics}
    \label{tab:hr_characteristics}
\end{table}
\\As shown in Figure~\ref{fig:hr_distribution}, the majority of the reviews were written between 2014 and 2018, with a sharp increase in 2016.
Figure~\ref{fig:emotions_hotel} provides an overview of the emotional composition, revealing that joy represents roughly 60\% of all instances, while neutral (20.2\%) and anger (10.4\%) follow as secondary categories.
Notably, this portion of the dataset differs substantially from the Telegram subset, both in temporal coverage and in emotional profile. While the hotel reviews are written in a joyful tone, mostly dating from 2014 to 2018, the Telegram messages instead, collected between June and August 2024, display a more neutral or even toxic tone, reflecting the  nature of online communication, especially when it happens in controversial topic.
Another difference between the two subsets is highlighted by the analysis of the distribution of text lengths: in Figure \ref{fig:text_length} the normalized density is plotted, allowing a comparison between the two dataset subset  regardless of corpus size. Note also that the x-axis is truncated at 400 words for better readability, as beyond this point densities fall below 0.002
for Hotel Reviews and 0.0002 for Telegram. This plot reveals significant differences between the two corpora under consideration: messages from Telegram exhibit a distribution strongly concentrated in the shorter length classes (0–40 words), indicating a tendency toward brevity and concise communication. In contrast, hotel reviews display a wider distribution with a long right tail, including texts extending to several hundred words. This variability suggests greater linguistic and argumentative complexity, consistent with the descriptive and evaluative nature of the reviews. The coexistence of these two distinct data sources, characterized by different emotional and stylistic features, enables us to evaluate how RAG systems perform across diverse linguistic contexts and communication styles.

\subsection{QA creation}
The process began by defining a set of ElasticSearch queries. For the Telegram subset, each query included filters on date, topic, and group chat. By iterating over these parameters and adding constraints on toxicity (Perplexity scores above 0.7~\cite{alvisi2025mappingitaliantelegramecosystem}), temporal range, and emotional tone, over 150,000 queries were generated. For the Hotel Reviews subset, filters were defined by combining city, postal code, review rating, emotion, and date interval, producing more than 100,000 combinations.\\
In both cases, the queries relied exclusively on metadata or ground-truth fields (e.g., topic, emotion). Only those returning more than 10 documents were retained to ensure sufficient complexity. This threshold allowed us to evaluate not only the system’s ability to generate accurate metadata-based filters but also its capacity to process and reason over richer informational contexts. Indeed, highly specific filters could yield only a single relevant document, offering the model an unambiguous yet trivial context. The 10-document criterion ensured a more challenging and meaningful evaluation. After filtering, we obtained 3,561 queries for Telegram and 3,064 for Hotel Reviews. For each query, the corresponding documents were stored in a dedicated file. Following an approach similar to~\citet{jeong2025videorag}, we used GPT-4o to generate three JSON entries per file containing a question, an answer, and the corresponding relevant document supporting the answer. However, not all generated triples were correct or relevant: some files yielded three valid QAs, others none. This process initially produced 10,683 QA pairs for Telegram and 9,192 for Hotel Reviews, a volume too large for full manual validation. To address this, GPT-4o was again used to select a single QA pair from each triplet, reducing the dataset to 2,999 QA pairs for Telegram and 1,780 for Hotel Reviews. These were then manually validated in two phases by seven annotators and four judges to ensure quality and reliability. Initially, the annotators evaluated each QA pair, either discarding or validating it based on the criteria reported in Appendix \ref{app:criteria}. Subsequently, the judges reviewed the QA pairs that had been validated by the annotators. This second round of evaluation aimed to ensure consistency and eliminate any residual errors or oversights. To avoid discarding too many QA pairs unnecessarily, annotators were allowed to modify the questions or answers (but not the original message or review) so that a pair that might otherwise be discarded could meet the positive evaluation criteria. For the Telegram dataset, out of the 2,999 initially annotated question–answer pairs, 1,187 were validated by the annotators, and 981 of these were also confirmed by the judges, resulting in a simple agreement of 82.6\%. For Hotel Reviews, 805 of the 1,780 initially annotated pairs were validated by the annotators, with 755 confirmed by the judges, corresponding to an agreement of 93.6\%. Overall, across the 4,779 initial pairs, 1,992 were validated by the annotators and 1,736 confirmed by the judges, yielding an overall agreement of 87.1\%. Finally, from the pool of validated QAs, we extracted a subset from each dataset subset and generated an additional set of questions whose answers were not contained in the text itself but derived from the associated metadata (e.g. date, city, rating, chat name). Since these new QAs were built upon previously validated items, they inherently satisfied the same quality and consistency constraints. This process produced 319 metadata-based QA pairs for Telegram and 545 for Hotel Reviews, leading to a balanced benchmark of 2,600 total QA pairs (1,300 per dataset subset). Consequently, the final benchmark comprises questions whose answers can either be inferred from the document text or are directly available in the metadata, ensuring a diverse and comprehensive evaluation setting.\\
By employing this rigorous two-stage validation process, the final benchmark consists exclusively of high-quality QA pairs. Furthermore, the benchmark supports the evaluation of systems on text understanding and metadata grounding.
\subsection{QA statistics}
\begin{figure}
  \centering
  \includegraphics[width=0.35\textwidth]{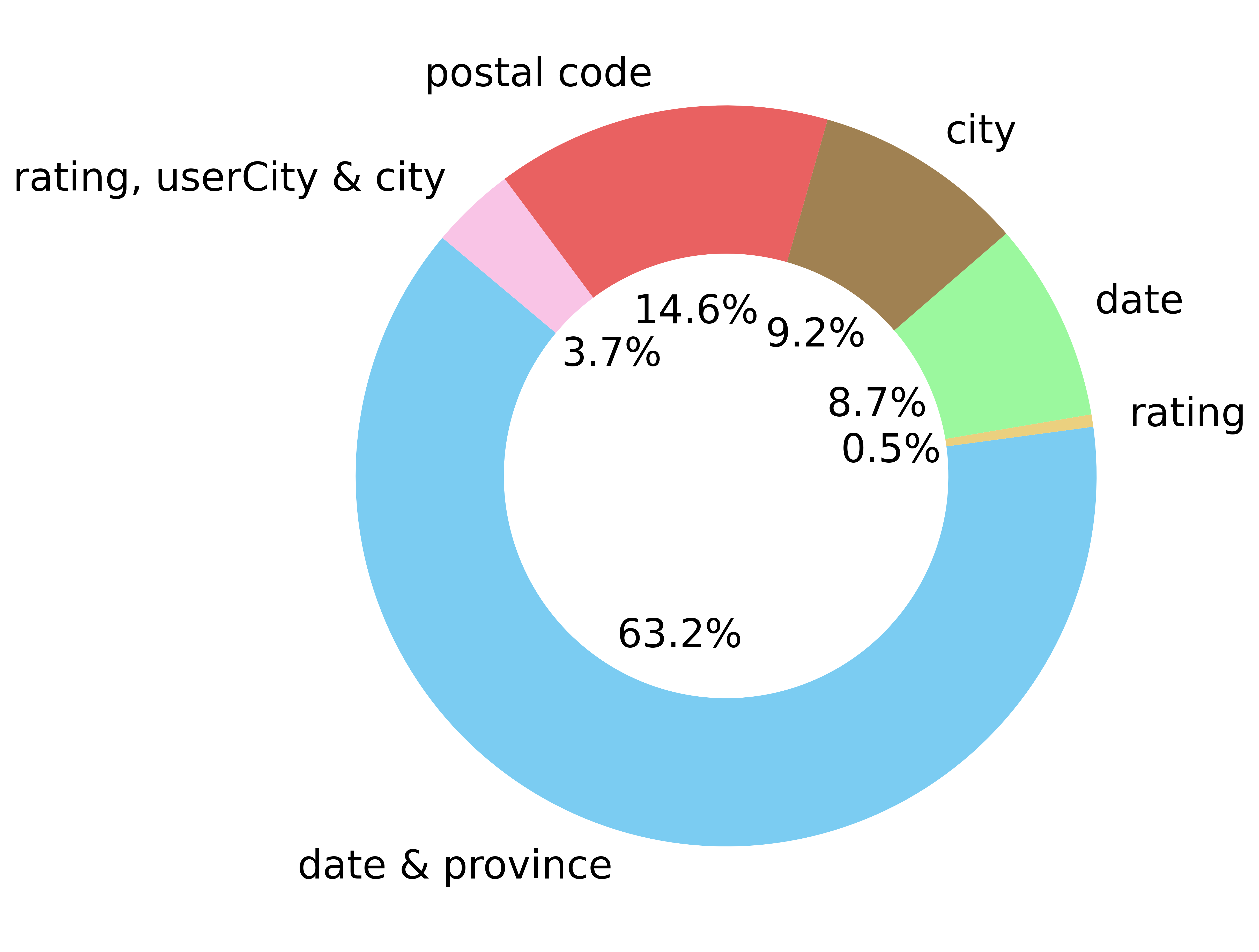}
  \caption{Distribution of the type of metadata constraints present in the hotel reviews Questions.}
  \label{fig:hotel_reviews_filters}
\end{figure}

As discussed in the previous section, the benchmark consists of a total of 2,600 QA pairs. Among them, approximately 33\% of the questions have answers corresponding to metadata fields, while the remaining 66\% have answers directly extractable from the text.\\
\begin{figure}
  \centering
  \includegraphics[width=0.35\textwidth]{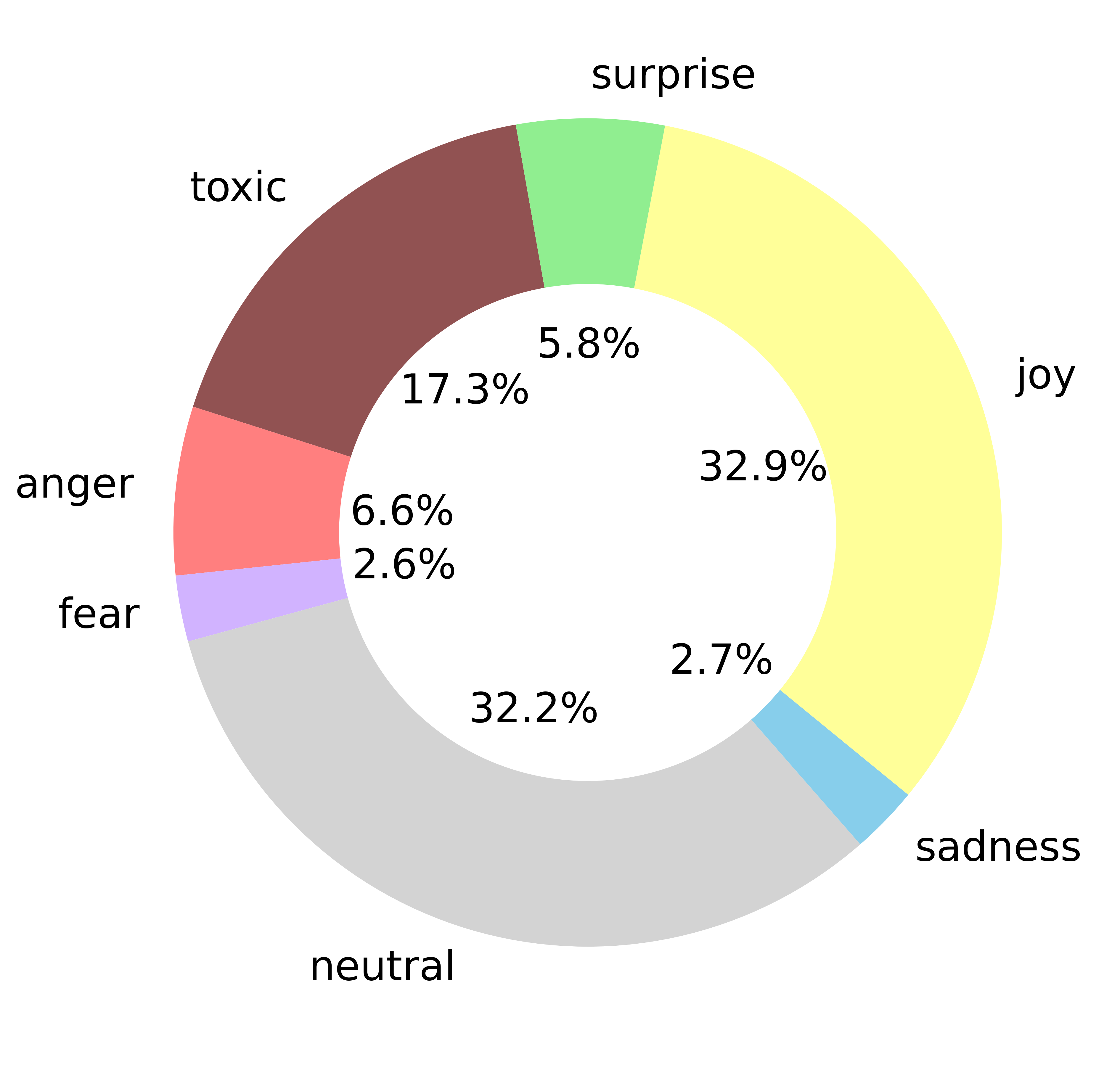}
  \caption{Distribution of tones across QA pairs.}
  \label{fig:distributions_emo_queries}
\end{figure}
Figure \ref{fig:distributions_emo_queries} illustrates the overall distribution of emotions and tones associated with the questions. However, when considering each subset individually, the distributions remain similar to those of the  dataset, as shown in Figures \ref{fig:distributions_c} and \ref{fig:emotions_hotel}. An exception emerges in the Telegram subset, where no questions are labeled with the anger emotion. This absence is due to the presence of the toxic label within the same subset: questions generated with emotion equal to anger constraint overlapped with those tagged as toxic, referring to identical metadata filters and relevant documents. To prevent redundancy, only the instances associated with the toxic tone were retained.\\
\begin{figure}
  \centering
  \includegraphics[width=0.48\textwidth]{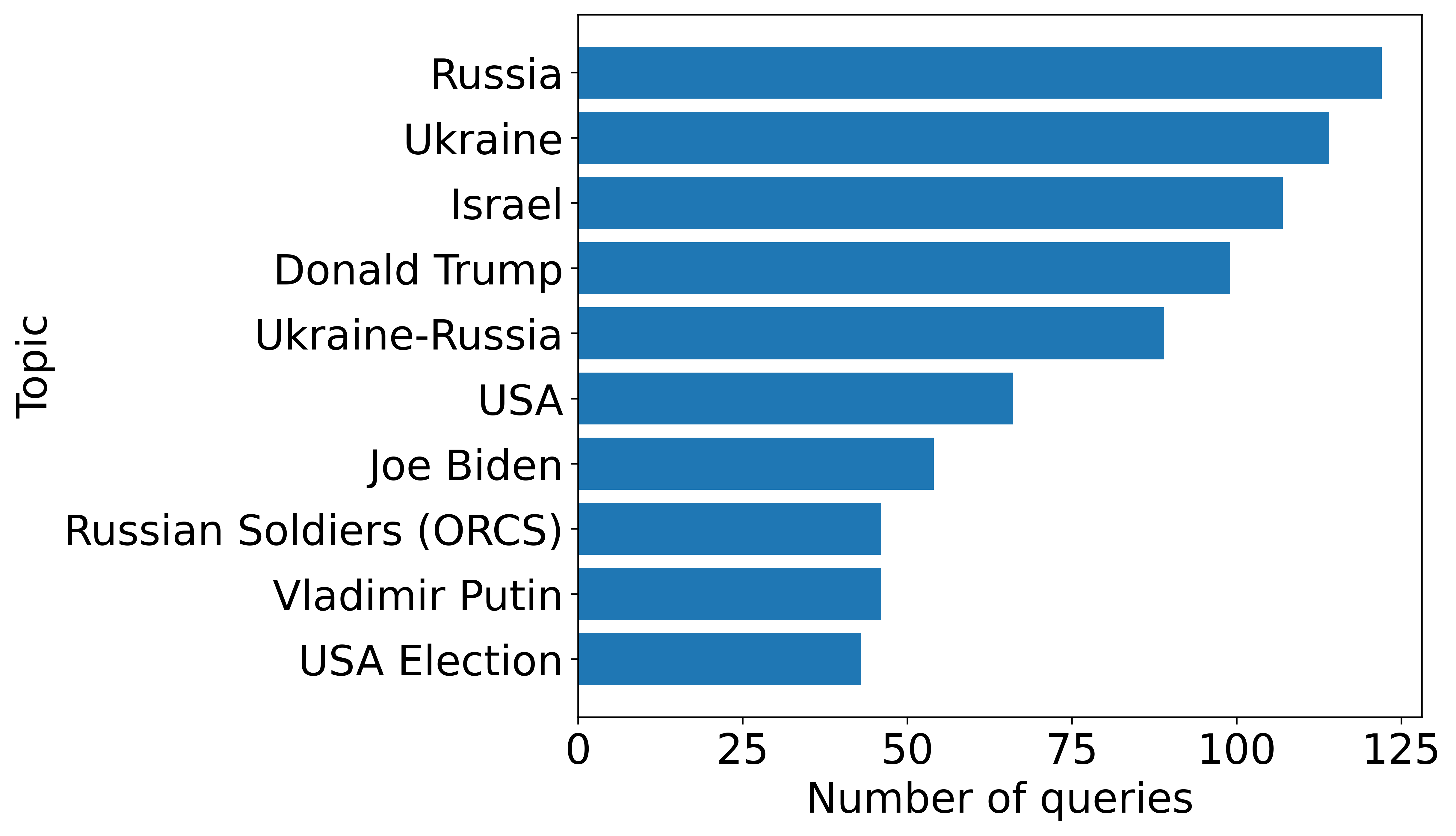}
  \caption{Distribution of the main topics present in the QAs. Only the top 10 topics are shown.}
  \label{fig:topic_queries}
\end{figure}
Figure \ref{fig:topic_queries} presents the topic distribution for the Telegram subset, which closely mirrors the thematic structure of the source dataset. Notably, the five most frequent topics in the questions correspond to the top topics in the dataset itself, albeit in a slightly different order. As depicted in the plot, the top 10 topics predominantly concern political figures and issues related to USA or the Russia–Ukraine conflict.\\
Concerning the distribution of metadata filters, questions in the Telegram subset include only constraints on the chat name and date. The temporal filter may be expressed as a specific day, as a range between two dates, or as before/after a certain date express in an implicit or explicit way. In contrast, the Hotel Reviews subset exhibits a wider variety of filter combinations, whose distributions are shown in Figure \ref{fig:hotel_reviews_filters}. Specifically, approximately 33\% of the questions apply a single filter on one metadata field, among postal code, city, date, or review rating, while around 63\% combine filters on date and province. The remaining 4\% of the questions include three simultaneous metadata filters, involving the review rating, the hotel’s city, and the reviewer’s city.

\subsection{Ethical consideration}\label{sec:ethical}
Our data collection was conducted with care to include only information from public Telegram channels and groups, explicitly excluding personal data such as usernames, phone numbers, user IDs, and chat IDs. The process aligns with Telegram’s Terms of Service~\cite{telegram_tos_2025} and Telegram API’s Terms of Service~\cite{telegram_api_terms_2025}, as neither prohibits the collection of public chat data. We have also taken care to avoid flooding the platform with requests during data collection with excessive requests. Furthermore, our research complies with the academic research exemptions outlined in Article 85 of the GDPR~\cite{gdpr_article85_2025}, which provide flexibility for processing publicly available data in the interest of freedom of expression and academic purposes. Since our work exclusively involves publicly accessible content and does not handle private user data, it qualifies for legitimate or public interest exemptions. Additionally, we follow the GDPR’s principle of data minimization by focusing solely on English textual content from public posts, ensuring that only the information necessary for our research objectives is processed~\cite{your_europe_gdpr}. Finally, as the dataset includes channels and groups that discuss controversial topics, it may contain some controversial messages.

\section{RAG systems evaluation methodology}\label{sec:rag_systems}
In this study, we employ the new AMAQA benchmark to evaluate RAG systems and address the following key research question:  
\begin{enumerate}
    \item[] \textbf{How does the inclusion of metadata affect the performance of RAG systems?}
\end{enumerate}

To answer to the research question, we implement and compare a set of complementary RAG architectures described below.

\textbf{Vanilla RAG.} This baseline follows the canonical RAG design~\cite{lewis2020retrieval}: the retriever encodes the input question into an embedding vector, compares it to a set of precomputed document embeddings using cosine similarity, and provide the top-$k$ most similar documents to the LLM which generates the answer based on them. Since the vanilla RAG cannot natively handle structured metadata alongside unstructured text, metadata are serialized into the textual content of each document before embedding.

\textbf{RAG with metadata filtering.} In this variant, the initial question, expressed in natural language, is translated, where possible, into a metadata filter  applicable to the data storage by an LLM. The filter restricts the search space. The remaining textual part of the question is then embedded and used for similarity retrieval among the filtered documents.

\textbf{\emph{Re²G}}~\cite{glass-etal-2022-re2g}. This variant augments the Vanilla RAG pipeline with a cross-encoder to re-score and reorder the retrieved passages before generation, testing whether neural reranking improves performance. This architecture allow us to evaluate whether improvements commonly observed from neural re-ranking extend to scenarios in which metadata have been embedded into the textual representation.

\textbf{\emph{Re²G*}} The most complete pipeline, combining explicit metadata filtering with reranking to explore their potential complementary effects.

By evaluating the performance of these architectures on the AMAQA benchmark, we aim to provide the community a baseline that is—to our knowledge—the first systematic comparison of RAG pipelines on a benchmark explicitly including metadata fields.

\subsection{Evaluation metrics} \label{sec:eval_metrics}
The evaluation of a RAG system depends on the specific task. In the case of Question Answering (QA), common metrics include Normalized Exact Match (NEM)\footnote{A variant of Exact Match that ignores formatting differences such as spaces, capitalization, and punctuation.} and the BLEU score~\cite{papineni2002bleu}. However, these metrics do not fully capture overall system performance, which is why it is standard practice to evaluate the retriever and generator components separately.

For the retriever, standard information retrieval metrics are used~\cite{gao2023retrieval}; in this work, we adopt the Mean Reciprocal Rank (MRR), as each query is associated with a single relevant document. For the responses generated by the LLM, traditional metrics based on textual or $n$-gram overlap fail to capture semantic quality. A more reliable approach would be human evaluation, although it is costly and time-consuming. As an alternative, LLM-as-a-Judge methods~\cite{zheng2023judging} provide a scalable solution: models such as GPT-4, trained through Reinforcement Learning from Human Feedback, have shown a high correlation with human judgments (over 80\%). In our study, we therefore use GPT-4.1 as the primary evaluator, computing accuracy based on its judgments. A sample of its evaluations was manually reviewed, and discrepancies were found in less than 1\% of cases, confirming its reliability. However, to maintain alignment with standard practice, we also report NEM as a complementary metric.

\subsection{Experimental setup}\label{sec:experimental_setup}
Experiments were conducted using the following configuration: document and query embeddings were obtained with \textit{BAAI/bge-base-en-v1.5}~\cite{xiaobaaiembeddings}, and compared via \textit{cosine similarity}. Retrieval was performed through exact \textit{k}-nearest neighbors (KNN) search. For experiments including metadata filtering, the retriever employed \textit{Mistral-nemo}~\cite{mistral-nemo} as the LLM. When a reranking stage was added, the \textit{BAAI/bge-reranker-large}~\cite{xiaobaaiembeddings} model was used as a cross-encoder. $\text{Re}^2\text{G}$ and $\text{Re}^2\text{G}$* experiments tested \textit{top\_n} values between 3 and 10, with \textit{k}=100 to preserve the \textit{k} $\gg$ \textit{top\_n} condition suggested by~\citet{yu2024rankrag}.\\ 

\section{Experimental Evaluation of RAG Systems with AMAQA}
This section evaluates the performance of RAG systems using the AMAQA dataset, focusing on the impact of metadata. %The results are analyzed in relation to the research questions introduced in Section~\ref{sec:rag_systems}.

\subsection{Evaluation of retriever performance}
\label{subsec:retriever_performance}
\begin{figure}[ht]
  \centering
  \includegraphics[width=0.45\textwidth]{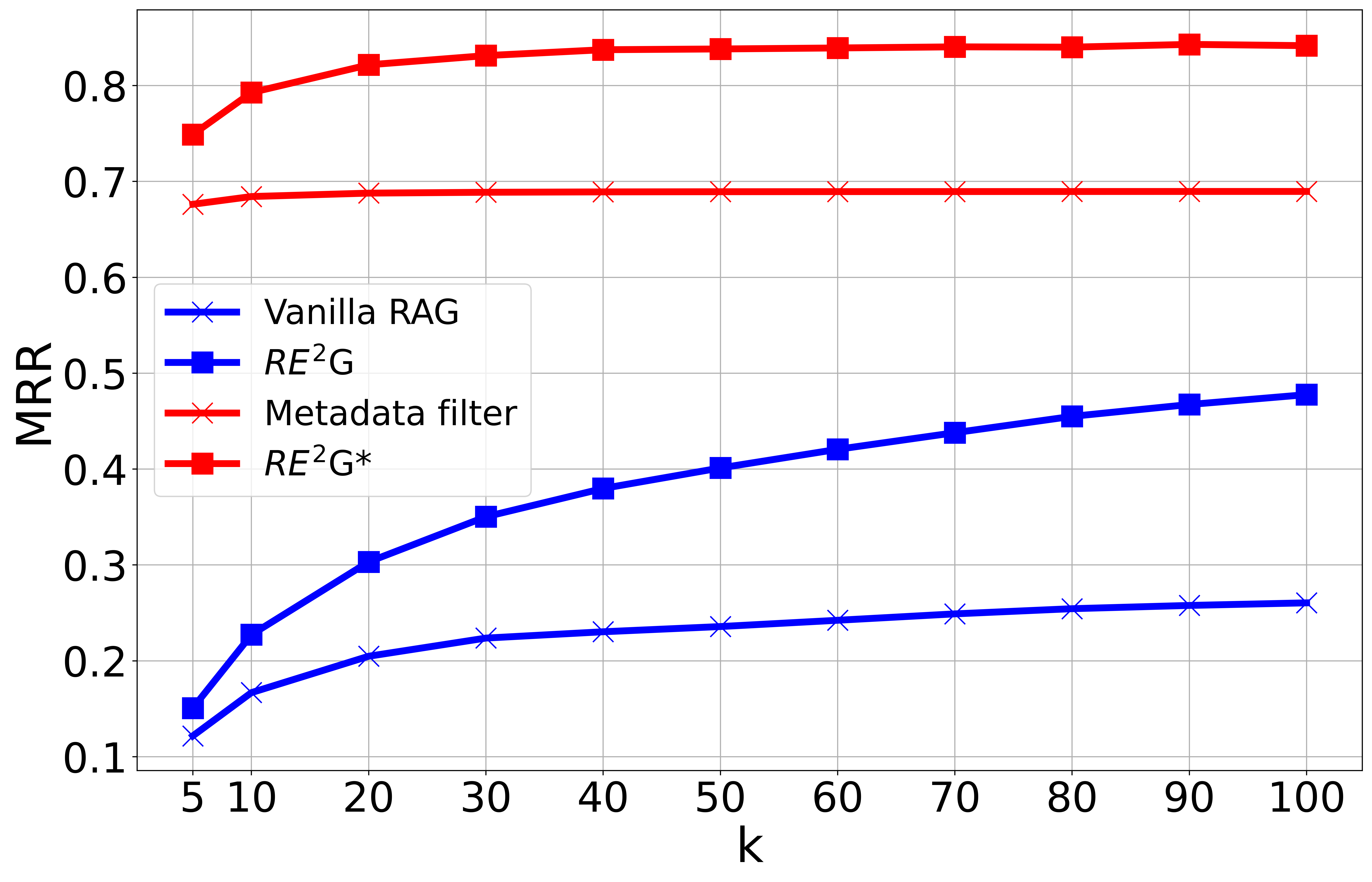}
  \caption{Mean Reciprocal Rank (MRR) as a function of k, comparing the performance of RAG architectures}
  \label{fig:mrr}
\end{figure}

\begin{figure*}
    \begin{subfigure}[b]{0.48\textwidth}
        \includegraphics[width=1.0\textwidth]{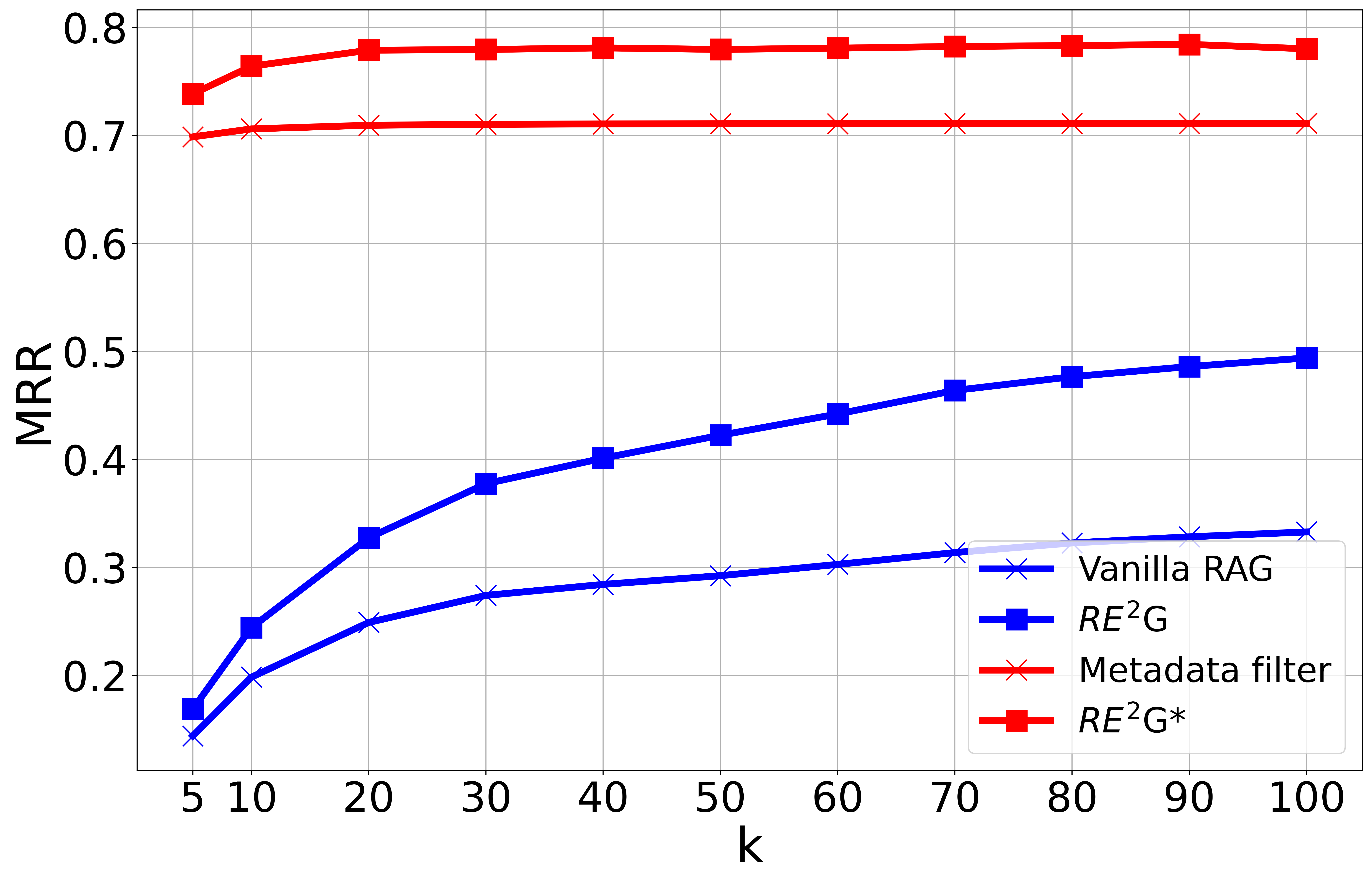}
  \caption{Mean Reciprocal Rank (MRR) comparison on Telegram messages' QA subset}
        \label{fig:MRR_AMAQA}
    \end{subfigure}
    \begin{subfigure}[b]{0.48\textwidth}
        \includegraphics[width=1.0\textwidth]{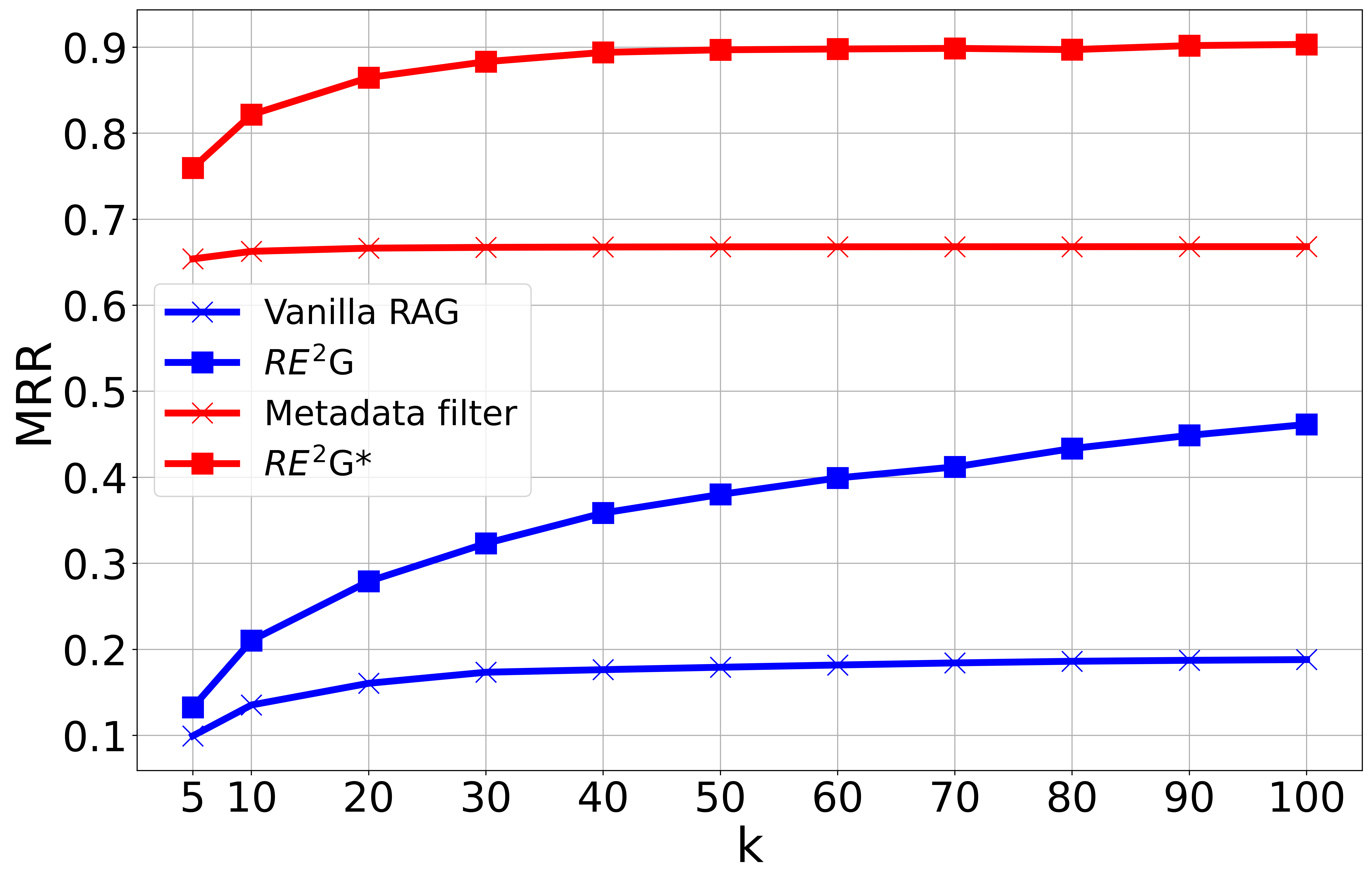}
          \caption{Mean Reciprocal Rank (MRR) comparison on Hotel Reviews' QA subset} 
        \label{fig:MRR_HOTEL}
    \end{subfigure}
    \caption{Comparison of MRR scores across different retrieval models on two QA subsets: Telegram messages (left) and Hotel Reviews (right).}
\end{figure*}

The analysis of the results related to the MRR (Mean Reciprocal Rank), as a function of the parameter number of retrieved documents $k$ in Figure \ref{fig:mrr} clearly highlights the critical impact of incorporating metadata filtering and reranking into the RAG framework when possible. The Vanilla RAG retriever shows the worst performance, reaching a maximum MRR of about 0.33 without reranker and 0.49 with it, even for the highest $k$ evaluated ($k$=100). This performance gap is attributable to the retriever architecture and the embedding strategy. Because metadata must be integrated into the text before embedding computation, the similarity calculation between embeddings fails to properly account for the query constraints. In practice, this serialization of metadata within the textual payload introduces semantic ambiguity: queries with strict filters (e.g., time ranges, cities, chat names) are compared to document vectors representing both message content and metadata, leading the retriever to overlook or dilute those constraints. The misalignment between semantic similarity and constraint satisfaction causes the retriever to rank the target document low, often returning texts that are semantically related but constraint-violating, or compliant yet irrelevant. The introduction of explicit metadata filtering proves essential. The LLM which transforms the constraints expressed in natural language into a filter that reduces and allows to consider only documents that respect those constraints drastically boosts performance: for a minimal $k$ of 5, the MRR rises from about 0.12 (Vanilla RAG) to 0.68 (RAG with metadata filter), a more than fourfold improvement that surpasses even the Vanilla RAG's retriever maximum performance even at K=100. The synergy between metadata filtering and reranking sets the highest benchmark, achieving an MRR of approximately 0.84 already at $k$=40 and remaining stable thereafter. The plot demonstrates that reranking improves performance to some extent, but the dominant contribution comes from effective metadata filtering. Moreover, a more granular analysis, shown in Figures \ref{fig:MRR_AMAQA} and \ref{fig:MRR_HOTEL}, reveals differences among data subsets, indicating that the higher complexity in the Telegram messages subset (maximum MRR 0.78) compared to the Hotel Reviews subset (maximum MRR 0.90) lies in the nature of the context. Queries over messages require more nuanced understanding, such as toxicity evaluation, and exhibit strong co-occurrence of messages with similar features, typical of social media (discussions focused over short time spans), which makes the retriever task more challenging even after effective metadata filtering.

\subsection{Evaluation of given answers}\label{subsec:vanilla-rag}
\begin{figure}[ht]
  \centering
  \includegraphics[width=0.48\textwidth]{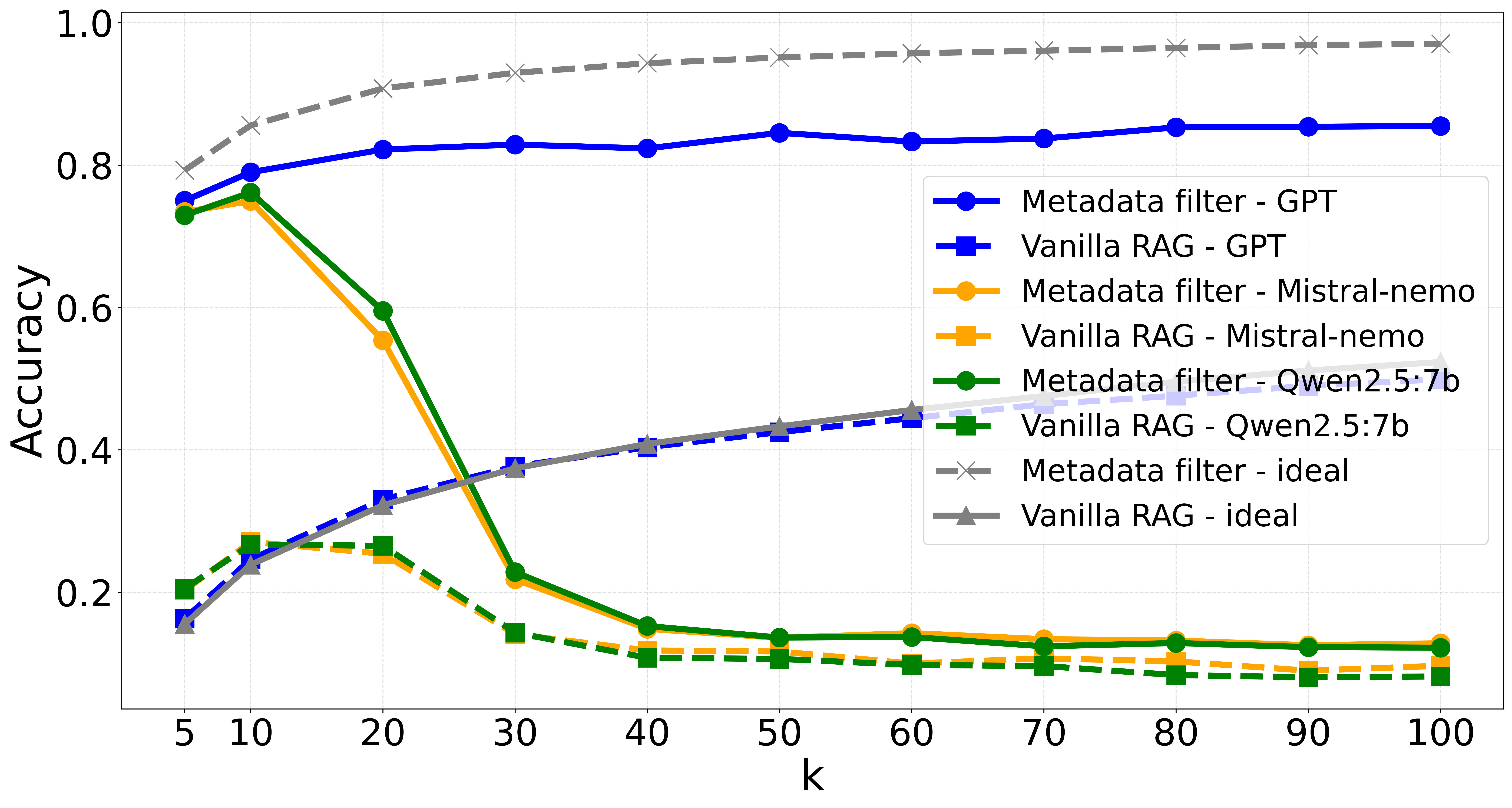}
  \caption{Accuracy as a function of k, comparing the performance of Vanilla RAG and RAG with metadata filtering} 
  \label{fig:total-acc}
\end{figure}
Figure \ref{fig:total-acc} shows the comparison of the accuracy obtained on the benchmark among Vanilla RAG and RAG with metadata filter. The plot also includes the ideal accuracy curves, representing the scenario in which, for each architecture, for every retrieved relevant document, the language model is able to extract the correct answer from it. It is important to note that accuracy is computed by comparing only the final answer generated by the RAG architecture with the ground-truth response. This approach tends to overestimate the performance of the vanilla version (as observed in some points where the accuracy even exceeds the ideal one), since, as discussed in Section \ref{sec:error}, in certain cases the model manages to extract the correct answer from non-relevant documents, specifically, from documents that are semantically related to the question but do not satisfy the constraints imposed by it (for example, when the retrieved messages fall outside the required temporal window). However, the plot in Figure \ref{fig:total-acc} reveals that, the use of RAG with metadata filtering shows superior performance compared to the Vanilla RAG configuration for lower $k$ values. This superiority can be attributed to the retriever’s effectiveness in identifying and retrieving a larger number of relevant documents when supported by filtering, as previously observed in the MRR metric analysis. However, for the open-source models tested (Qwen2.5:7b~\cite{qwen2025qwen25technicalreport} and Mistral-nemo), the introduction of a large number of retrieved documents quickly nullifies this advantage. Starting from K>20, there is a sharp drop in accuracy for the metadata-filtered configurations, bringing them in line with the performance of their respective vanilla versions. This phenomenon is due to the well known Lost-in-the-Middle problem~\cite{liu2024lost}, which indicates that smaller LLMs struggle to focus on relevant information when it is buried within a very large context, effectively negating all the benefits of the introduction of metadata filtering. In particular, Figure~\ref{fig:accuracy_tg} shows that this effect emerges in the Telegram Messages QA task for $k \geq 30$, while Figure~\ref{fig:accuracy_hr} highlights a similar trend in the Hotel Reviews dataset for $k \geq 20$. This confirms that performance degradation occurs earlier for the Hotel Reviews task, as the average text length is greater, as illustrated in Figure~\ref{fig:text_length}. In contrast, the GPT-4o model maintains high accuracy even for larger $k$ values. Its ability to effectively process extended contexts reinforces the advantage of the metadata-filtered retriever, ensuring consistently superior performance compared to Vanilla RAG. 
\begin{figure}[ht]
  \centering
  \includegraphics[width=0.48\textwidth]{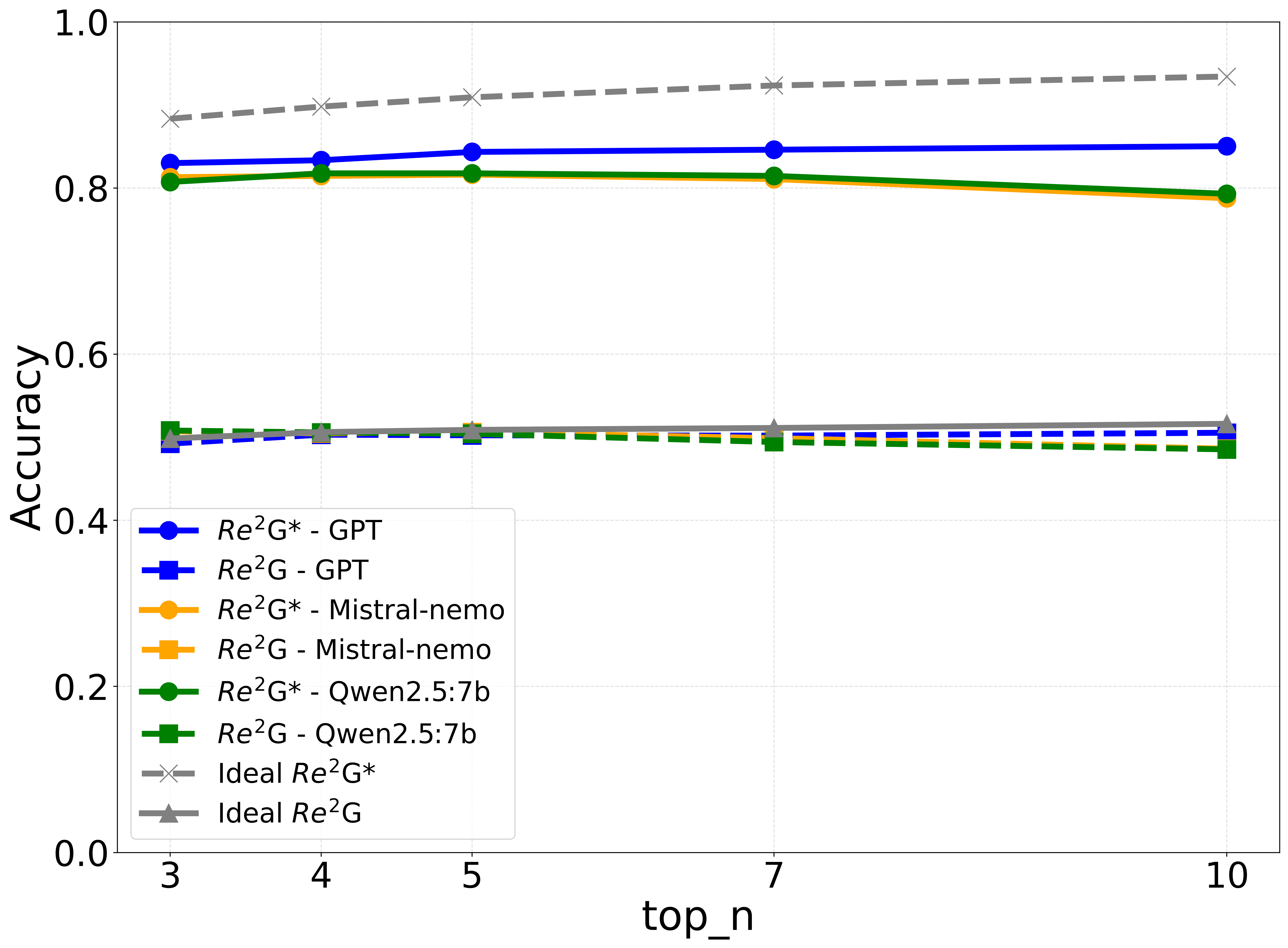}
  \caption{Accuracies as function of k, comparing the performance of two architectures involving a reranker phase.} 
  \label{fig:total-acc-reranker}
\end{figure}
As shown in Figure \ref{fig:total-acc-reranker} the integration of a reranker into both RAG architectures led to an improvement in accuracy, although not a drastically significant one. The adoption of the reranker enabled the open-source models to almost match the highest performance levels achieved by GPT-4o with ($k$ = 100), while attaining these results with a much smaller number of input documents (top\_n between $( 3 \le n \le 10 ))$.\\
All these observations are also confirmed by the behavior of the NEM metric, which exhibits a similar trend. The only additional remark concerns GPT-4o, whose NEM values for Re²G* evaluations are slightly lower than that of the tested open-source models, as we can see from Figures \ref{fig:nem_1} and \ref{fig:nem_2} in Appendix \ref{app:c}.

\subsection{Anecdotal Error Analysis} \label{sec:error}
By analyzing a sample of incorrectly generated responses, we identified patterns in RAG architectures that highlight critical interactions between the retrieval and generation phases. In Vanilla RAG and Re²G, we observed cases where the model provided the correct answer even though the retriever hadn't identified the relevant document. This occurred when another retrieved text expressed the same concept or provided semantically equivalent information, even though it didn't satisfy the metadata constraints specified in the query. This situation occurred both when the answer had to be extracted from the text and when it matched a metadata value (e.g., number of reactions or ratings). In such cases, documents with similar content or terminology, but with metadata inconsistent with the constraints, still coincidentally reported the required values, thus producing the correct answer.\\
However for RAG with metadata filtering and Re²G* architectures, we still observed several sources of error. First, the LLM occasionally failed to recognize the presence of the relevant document, incorrectly responding that the required information was not available in the provided context. This issue, however, did not occur in out tests when a reranker was employed, likely because the relevant documents were promoted to the top positions (thus mitigating the lost-in-the-middle problem) and because passing only 3 to 10 documents reduced the overall context length. Second, some errors originated upstream at the retrieval stage: even if metadata constraints were always fulfilled, the relevant document was sometimes missing from the retrieved set, preventing the model from generating the correct answer. Finally, certain failures arose when both the relevant document and semantically similar, but non-relevant documents were retrieved, indicating that the model struggled to differentiate between distractors and the truly relevant source.
\subsection{Summary of Experimental Findings}
\label{subsec:summary_results}
Table~\ref{table:results} presents a comprehensive summary of the performance of the evaluated RAG architectures. Note that for open-source LLMs, we report the results for Qwen2.5:7B; however, the results obtained with Mistral-Nemo are nearly identical and show no statistically significant differences. Additional details on results obtained for each QA subset are provided in Table~\ref{tab:total_res} in Appendix~\ref{app:c}.

\begin{table}[ht]
\resizebox{0.48\textwidth}{!}{%
\begin{tabular}{l|c|c|c|c|c}
\toprule
\textit{RAG architecture} &  \textit{LLM} & \textit{Acc.} & \textit{NEM} & \textit{k} & \textit{top\_n} \\
\midrule
$\text{Re}^2\text{G}$* & GPT-4o & \textbf{0.85} & \textbf{0.63}  & 100 & 10 \\
Metadata filtering & GPT-4o & \textbf{0.86} & \textbf{0.63}  & 100 & - \\
$\text{Re}^2\text{G}$ & GPT-4o & 0.51 & 0.37 & 100 & 10 \\
Vanilla RAG & GPT-4o & 0.50 & 0.37 & 100 & - \\
\midrule
$\text{Re}^2\text{G}$* & Qwen2.5:7b & 0.82 & 0.65  & 100 & 3 \\
Metadata filtering & Qwen2.5:7b & 0.76 & 0.61  & 10 & - \\
$\text{Re}^2\text{G}$ & Qwen2.5:7b & 0.51 & 0.39 & 100 & 5 \\
Vanilla RAG & Qwen2.5:7b & 0.27 & 0.20 & 10 & - \\
\bottomrule
\end{tabular}
}
\caption{Comparison of RAG architectures under optimal settings based on accuracy (acc.) and NEM. Bold marks the \textbf{best} result.}
\label{table:results}
\end{table}

Regarding the Research Question "\textit{How does the inclusion of metadata affect the performance of RAG systems?}", the experimental results demonstrate that metadata filtering plays a crucial role in enhancing retrieval precision, leading to an increase in accuracy from 0.27 to 0.76 for open-source models and from 0.5 and 0.86 for GPT-4o\footnote{Note that there aren't statistically differences between $\text{Re}^2\text{G}$ and Metadata filtering architecture when the LLM used is GPT-4o}. Furthermore, our experiments showed that well-known techniques, such as the use of rerankers, improve performance: in our tests, open-source models increase the accuracy value from 0.76 to 0.82. However, in scenarios where questions impose constraints on metadata, it is the management of these metadata that has the greatest impact on the results. These experimental results provide a robust foundation for future advancements in RAG research for metadata-driven question answering systems.

\section{Limitations}\label{secc:limitation}
This study presents some limitations that should be acknowledged. First, the metadata schema used in the experiments is not overly complex and limited in size (up to eleven distinct fields per subset). It remains unclear whether the retriever, and in particular the LLM component responsible for generating the structured query from the natural-language question, would maintain comparable performance when scaled to larger or more complex metadata, because each field requires a short textual description and illustrative examples within the prompt to guide filter generation. Although no performance degradation was observed across tested LLMs (and thus omitted from the main discussion for brevity), this finding may not generalize to richer or more heterogeneous metadata settings. Moreover, the metadata used are semantically distinct from one another: for instance, \textit{City} and \textit{Rating} are considerably less related than attributes such as \textit{userName} and \textit{userScreenName} (which were excluded to be complaint with GDPR constraints, as discussed in Section \ref{sec:ethical}, but may be relevant in real-world deployments). Consequently, the approach has not been evaluated on highly semantically similar metadata, which could influence the LLM’s disambiguation capability during structured query generation. Another limitation concerns metadata normalization. The current approach assumes a predefined set of keys and standardized values. For example, filtering by city equal to "New York City" would not retrieve entries labeled as "NY" or "New York". Similarly, entries using a different key, such as "location" instead of "city", would also be excluded, even if the value is identical. Additionally, the benchmark employed targets only single-hop question answering, i.e., cases in which the relevant information can be retrieved from a single source. Finally, certain practical and model-related factors may have affected the system’s performance. Azure’s API filter for harmful content could not be disabled, potentially having a mildly negative impact on GPT-4o’s results. Furthermore, the system’s effectiveness is inherently tied to the capacity of the LLMs employed, none exceeding 10 billion parameters in the open-source configuration, thus constraining the expressiveness of the Generator. The results also exhibit sensitivity to prompt design, particularly with respect to context formatting. It is reasonable to expect that employing more capable models, in combination with more extensive prompt engineering, could further enhance performance.

\section{Conclusion and Future Work}
In this work, we introduced AMAQA, the first open-access QA dataset specifically designed to evaluate Retrieval-Augmented Generation (RAG) systems that leverage metadata. AMAQA, built on 20 thousands Hotel Reviews and 1.1 million messages from 26 public Telegram groups and enriched with structured metadata, enables sophisticated QA tasks and offers a novel resource for advancing metadata-driven systems. Through our experimentation, we demonstrated that metadata filtering significantly improves retrieval accuracy, boosting system accuracy from 0.27 to 0.76 for open-source models and from 0.5 to 0.86 for GPT-4o. Furthermore, by integrating a reranker, we were able to further increase accuracy up to 0.82 for open-source models while reducing input token usage across all models. Nevertheless, our findings highlight that the most critical factor for improving RAG performance remains the effective management and exploitation of metadata when available.

Future work could achieve significant improvements by fine-tuning the embeddings, which could lead to better performance in the retrieval phase or, alternatively, fine-tuning the LLM used as a generator could ensure that the responses are more aligned with the specific terms and language present in the data~\cite{gao2023retrieval}. Moreover, it would be valuable to investigate potential limitations related to the number, type, and semantic meaning of metadata. In addition to methodological refinements, it is equally important to maintain the benchmark over time, to prevent obsolescence and expand or adapt the knowledge base to other domains. Furthermore, the benchmark could be extended to include multi-hop question answering, enabling the evaluation of more complex reasoning and retrieval chains. 

By establishing AMAQA as a benchmark and demonstrating the impact of metadata-driven retrieval techniques, we aim to inspire further research into more context-aware, robust, and efficient QA systems. We make the dataset publicly available, in an anonymous public repository, to encourage collaboration and future advancements in this field.

\bibliography{biblio}

@inproceedings{baumgartner2020pushshift,
  title={The pushshift telegram dataset},
  author={Baumgartner, Jason and Zannettou, Savvas and Squire, Megan and Blackburn, Jeremy},
  booktitle={Proceedings of the international AAAI conference on web and social media},
  volume={14},
  pages={840--847},
  year={2020}
}

@article{hou2024bridging,
  title={Bridging language and items for retrieval and recommendation},
  author={Hou, Yupeng and Li, Jiacheng and He, Zhankui and Yan, An and Chen, Xiusi and McAuley, Julian},
  journal={arXiv preprint arXiv:2403.03952},
  year={2024}
}

@inproceedings{10.1145/3690624.3709397,
author = {La Morgia, Massimo and Mei, Alessandro and Mongardini, Alberto Maria},
title = {TGDataset: Collecting and Exploring the Largest Telegram Channels Dataset},
year = {2025},
isbn = {9798400712456},
publisher = {Association for Computing Machinery},
address = {New York, NY, USA},
url = {https://doi.org/10.1145/3690624.3709397},
doi = {10.1145/3690624.3709397},
booktitle = {Proceedings of the 31st ACM SIGKDD Conference on Knowledge Discovery and Data Mining V.1},
pages = {2325–2334},
numpages = {10},
location = {Toronto ON, Canada},
series = {KDD '25}
}

@misc{banik_movies_dataset,
  author = {Rounak Banik},
  title = {The Movies Dataset},
  year = {2017},
  url = {https://www.kaggle.com/datasets/rounakbanik/the-movies-dataset}
}

@inproceedings{yang-etal-2018-hotpotqa,
    title = "{H}otpot{QA}: A Dataset for Diverse, Explainable Multi-hop Question Answering",
    author = "Yang, Zhilin  and
      Qi, Peng  and
      Zhang, Saizheng  and
      Bengio, Yoshua  and
      Cohen, William  and
      Salakhutdinov, Ruslan  and
      Manning, Christopher D.",
    editor = "Riloff, Ellen  and
      Chiang, David  and
      Hockenmaier, Julia  and
      Tsujii, Jun{'}ichi",
    booktitle = "Proceedings of the 2018 Conference on Empirical Methods in Natural Language Processing",
    month = oct # "-" # nov,
    year = "2018",
    address = "Brussels, Belgium",
    publisher = "Association for Computational Linguistics",
    url = "https://aclanthology.org/D18-1259/",
    doi = "10.18653/v1/D18-1259",
    pages = "2369--2380"
}

@inproceedings{rein2024gpqa,
      title={{GPQA}: A Graduate-Level Google-Proof Q\&A Benchmark},
      author={David Rein and Betty Li Hou and Asa Cooper Stickland and Jackson Petty and Richard Yuanzhe Pang and Julien Dirani and Julian Michael and Samuel R. Bowman},
      booktitle={First Conference on Language Modeling},
      year={2024},
      url={https://openreview.net/forum?id=Ti67584b98}
}

@article{reddy2019coqa,
  title={Coqa: A conversational question answering challenge},
  author={Reddy, Siva and Chen, Danqi and Manning, Christopher D},
  journal={Transactions of the Association for Computational Linguistics},
  volume={7},
  pages={249--266},
  year={2019},
  publisher={MIT Press One Rogers Street, Cambridge, MA 02142-1209, USA journals-info~…}
}

@article{gao2023retrieval,
  title={Retrieval-augmented generation for large language models: A survey},
  author={Gao, Yunfan and Xiong, Yun and Gao, Xinyu and Jia, Kangxiang and Pan, Jinliu and Bi, Yuxi and Dai, Yi and Sun, Jiawei and Wang, Haofen},
  journal={arXiv preprint arXiv:2312.10997},
  year={2023}
}

@inproceedings{glass-etal-2022-re2g,
    title = "{R}e2{G}: Retrieve, Rerank, Generate",
    author = "Glass, Michael  and
      Rossiello, Gaetano  and
      Chowdhury, Md Faisal Mahbub  and
      Naik, Ankita  and
      Cai, Pengshan  and
      Gliozzo, Alfio",
    editor = "Carpuat, Marine  and
      de Marneffe, Marie-Catherine  and
      Meza Ruiz, Ivan Vladimir",
    booktitle = "Proceedings of the 2022 Conference of the North American Chapter of the Association for Computational Linguistics: Human Language Technologies",
    month = jul,
    year = "2022",
    address = "Seattle, United States",
    publisher = "Association for Computational Linguistics",
    url = "https://aclanthology.org/2022.naacl-main.194/",
    doi = "10.18653/v1/2022.naacl-main.194",
    pages = "2701--2715"
}

@article{ekman1992argument,
  title={An argument for basic emotions},
  author={Ekman, Paul},
  journal={Cognition \& emotion},
  volume={6},
  number={3-4},
  pages={169--200},
  year={1992},
  publisher={Taylor \& Francis}
}

@article{seyeditabari2018emotion,
  title={Emotion detection in text: a review},
  author={Seyeditabari, Armin and Tabari, Narges and Zadrozny, Wlodek},
  journal={arXiv preprint arXiv:1806.00674},
  year={2018}
}

@inproceedings{plaza-del-arco-etal-2022-natural,
    title = "Natural Language Inference Prompts for Zero-shot Emotion Classification in Text across Corpora",
    author = "Plaza-del-Arco, Flor Miriam  and
      Mart{\'i}n-Valdivia, Mar{\'i}a-Teresa  and
      Klinger, Roman",
    editor = "Calzolari, Nicoletta  and
      Huang, Chu-Ren  and
      Kim, Hansaem  and
      Pustejovsky, James  and
      Wanner, Leo  and
      Choi, Key-Sun  and
      Ryu, Pum-Mo  and
      Chen, Hsin-Hsi  and
      Donatelli, Lucia  and
      Ji, Heng  and
      Kurohashi, Sadao  and
      Paggio, Patrizia  and
      Xue, Nianwen  and
      Kim, Seokhwan  and
      Hahm, Younggyun  and
      He, Zhong  and
      Lee, Tony Kyungil  and
      Santus, Enrico  and
      Bond, Francis  and
      Na, Seung-Hoon",
    booktitle = "Proceedings of the 29th International Conference on Computational Linguistics",
    month = oct,
    year = "2022",
    address = "Gyeongju, Republic of Korea",
    publisher = "International Committee on Computational Linguistics",
    url = "https://aclanthology.org/2022.coling-1.592/",
    pages = "6805--6817"
}

@article{laurer_less_2023,
    title = {Less {Annotating}, {More} {Classifying}: {Addressing} the {Data} {Scarcity} {Issue} of {Supervised} {Machine} {Learning} with {Deep} {Transfer} {Learning} and {BERT}-{NLI}},
    issn = {1047-1987, 1476-4989},
    shorttitle = {Less {Annotating}, {More} {Classifying}},
    url = {https://www.cambridge.org/core/product/identifier/S1047198723000207/type/journal_article},
    doi = {10.1017/pan.2023.20},
    language = {en},
    urldate = {2023-06-20},
    journal = {Political Analysis},
    author = {Laurer, Moritz and Van Atteveldt, Wouter and Casas, Andreu and Welbers, Kasper},
    month = jun,
    year = {2023},
    pages = {1--33},
}

@article{zheng2023judging,
  title={Judging llm-as-a-judge with mt-bench and chatbot arena},
  author={Zheng, Lianmin and Chiang, Wei-Lin and Sheng, Ying and Zhuang, Siyuan and Wu, Zhanghao and Zhuang, Yonghao and Lin, Zi and Li, Zhuohan and Li, Dacheng and Xing, Eric and others},
  journal={Advances in Neural Information Processing Systems},
  volume={36},
  pages={46595--46623},
  year={2023}
}

@article{lyu2024crud,
  title={Crud-rag: A comprehensive chinese benchmark for retrieval-augmented generation of large language models},
  author={Lyu, Yuanjie and Li, Zhiyu and Niu, Simin and Xiong, Feiyu and Tang, Bo and Wang, Wenjin and Wu, Hao and Liu, Huanyong and Xu, Tong and Chen, Enhong},
  journal={ACM Transactions on Information Systems},
  year={2024},
  publisher={ACM New York, NY}
}

@inproceedings{
yu2024rankrag,
title={Rank{RAG}: Unifying Context Ranking with Retrieval-Augmented Generation in {LLM}s},
author={Yue Yu and Wei Ping and Zihan Liu and Boxin Wang and Jiaxuan You and Chao Zhang and Mohammad Shoeybi and Bryan Catanzaro},
booktitle={The Thirty-eighth Annual Conference on Neural Information Processing Systems},
year={2024},
url={https://openreview.net/forum?id=S1fc92uemC}
}

@misc{your_europe_gdpr,
  author       = "{Your Europe}",
  title        = "{Data protection under GDPR}",
  year         = {2025},
  month        = {January},
  day          = {27},
  url          = {https://europa.eu/youreurope/business/dealing-with-customers/dataprotection/data-protection-gdpr/index_en.htm},
  note         = {Accessed: 2025-01-27}
}

@inproceedings{wang-etal-2023-text2topic,
    title = "{T}ext2{T}opic: Multi-Label Text Classification System for Efficient Topic Detection in User Generated Content with Zero-Shot Capabilities",
    author = "Wang, Fengjun  and
      Beladev, Moran  and
      Kleinfeld, Ofri  and
      Frayerman, Elina  and
      Shachar, Tal  and
      Fainman, Eran  and
      Lastmann Assaraf, Karen  and
      Mizrachi, Sarai  and
      Wang, Benjamin",
    editor = "Wang, Mingxuan  and
      Zitouni, Imed",
    booktitle = "Proceedings of the 2023 Conference on Empirical Methods in Natural Language Processing: Industry Track",
    month = dec,
    year = "2023",
    address = "Singapore",
    publisher = "Association for Computational Linguistics",
    url = "https://aclanthology.org/2023.emnlp-industry.10/",
    doi = "10.18653/v1/2023.emnlp-industry.10",
    pages = "93--103"
}

@article{grootendorst2022bertopic,
  title={BERTopic: Neural topic modeling with a class-based TF-IDF procedure},
  author={Grootendorst, Maarten},
  journal={arXiv preprint arXiv:2203.05794},
  year={2022}
}

@article{zhou2023instruction,
  title={Instruction-following evaluation for large language models},
  author={Zhou, Jeffrey and Lu, Tianjian and Mishra, Swaroop and Brahma, Siddhartha and Basu, Sujoy and Luan, Yi and Zhou, Denny and Hou, Le},
  journal={arXiv preprint arXiv:2311.07911},
  year={2023}
}

@article{liu2024lost,
  title={Lost in the middle: How language models use long contexts},
  author={Liu, Nelson F and Lin, Kevin and Hewitt, John and Paranjape, Ashwin and Bevilacqua, Michele and Petroni, Fabio and Liang, Percy},
  journal={Transactions of the Association for Computational Linguistics},
  volume={12},
  pages={157--173},
  year={2024},
  publisher={MIT Press One Broadway, 12th Floor, Cambridge, Massachusetts 02142, USA~…}
}

@article{nguyen2016ms,
  title={Ms marco: A human-generated machine reading comprehension dataset},
  author={Nguyen, Tri and Rosenberg, Mir and Song, Xia and Gao, Jianfeng and Tiwary, Saurabh and Majumder, Rangan and Deng, Li},
  year={2016}
}

@inproceedings{xiong-etal-2024-benchmarking,
    title = "Benchmarking Retrieval-Augmented Generation for Medicine",
    author = "Xiong, Guangzhi  and
      Jin, Qiao  and
      Lu, Zhiyong  and
      Zhang, Aidong",
    editor = "Ku, Lun-Wei  and
      Martins, Andre  and
      Srikumar, Vivek",
    booktitle = "Findings of the Association for Computational Linguistics: ACL 2024",
    month = aug,
    year = "2024",
    address = "Bangkok, Thailand",
    publisher = "Association for Computational Linguistics",
    url = "https://aclanthology.org/2024.findings-acl.372/",
    doi = "10.18653/v1/2024.findings-acl.372",
    pages = "6233--6251"
}

@inproceedings{mallen-etal-2023-trust,
    title = "When Not to Trust Language Models: Investigating Effectiveness of Parametric and Non-Parametric Memories",
    author = "Mallen, Alex  and
      Asai, Akari  and
      Zhong, Victor  and
      Das, Rajarshi  and
      Khashabi, Daniel  and
      Hajishirzi, Hannaneh",
    editor = "Rogers, Anna  and
      Boyd-Graber, Jordan  and
      Okazaki, Naoaki",
    booktitle = "Proceedings of the 61st Annual Meeting of the Association for Computational Linguistics (Volume 1: Long Papers)",
    month = jul,
    year = "2023",
    address = "Toronto, Canada",
    publisher = "Association for Computational Linguistics",
    url = "https://aclanthology.org/2023.acl-long.546/",
    doi = "10.18653/v1/2023.acl-long.546",
    pages = "9802--9822"
}

@inproceedings{joshi-etal-2017-triviaqa,
    title = "{T}rivia{QA}: A Large Scale Distantly Supervised Challenge Dataset for Reading Comprehension",
    author = "Joshi, Mandar  and
      Choi, Eunsol  and
      Weld, Daniel  and
      Zettlemoyer, Luke",
    editor = "Barzilay, Regina  and
      Kan, Min-Yen",
    booktitle = "Proceedings of the 55th Annual Meeting of the Association for Computational Linguistics (Volume 1: Long Papers)",
    month = jul,
    year = "2017",
    address = "Vancouver, Canada",
    publisher = "Association for Computational Linguistics",
    url = "https://aclanthology.org/P17-1147/",
    doi = "10.18653/v1/P17-1147",
    pages = "1601--1611"
}

@inproceedings{rajpurkar-etal-2016-squad,
    title = "{SQ}u{AD}: 100,000+ Questions for Machine Comprehension of Text",
    author = "Rajpurkar, Pranav  and
      Zhang, Jian  and
      Lopyrev, Konstantin  and
      Liang, Percy",
    editor = "Su, Jian  and
      Duh, Kevin  and
      Carreras, Xavier",
    booktitle = "Proceedings of the 2016 Conference on Empirical Methods in Natural Language Processing",
    month = nov,
    year = "2016",
    address = "Austin, Texas",
    publisher = "Association for Computational Linguistics",
    url = "https://aclanthology.org/D16-1264",
    doi = "10.18653/v1/D16-1264",
    pages = "2383--2392",
    eprint={1606.05250},
    archivePrefix={arXiv},
    primaryClass={cs.CL},
}

@inproceedings{perez2018you,
  title={You are your metadata: Identification and obfuscation of social media users using metadata information},
  author={Perez, Beatrice and Musolesi, Mirco and Stringhini, Gianluca},
  booktitle={Proceedings of the International AAAI Conference on Web and Social Media},
  volume={12},
  number={1},
  year={2018}
}

@article{pelofske2023cybersecurity,
  title={Cybersecurity Threat hunting and vulnerability analysis using a Neo4j graph database of open source intelligence},
  author={Pelofske, Elijah and Liebrock, Lorie M and Urias, Vincent},
  journal={arXiv preprint arXiv:2301.12013},
  year={2023}
}

@article{jeong2025videorag,
  title={VideoRAG: Retrieval-Augmented Generation over Video Corpus},
  author={Jeong, Soyeong and Kim, Kangsan and Baek, Jinheon and Hwang, Sung Ju},
  journal={arXiv preprint arXiv:2501.05874},
  year={2025}
}

@article{gilardi2023chatgpt,
  title={ChatGPT outperforms crowd workers for text-annotation tasks},
  author={Gilardi, Fabrizio and Alizadeh, Meysam and Kubli, Ma{\"e}l},
  journal={Proceedings of the National Academy of Sciences},
  volume={120},
  number={30},
  pages={e2305016120},
  year={2023},
  publisher={National Academy of Sciences}
}

@article{munker2024zero,
  title={Zero-shot prompt-based classification: topic labeling in times of foundation models in German Tweets},
  author={M{\"u}nker, Simon and Kugler, Kai and Rettinger, Achim},
  journal={arXiv preprint arXiv:2406.18239},
  year={2024}
}

@inproceedings{mu-etal-2024-large,
    title = "Large Language Models Offer an Alternative to the Traditional Approach of Topic Modelling",
    author = "Mu, Yida  and
      Dong, Chun  and
      Bontcheva, Kalina  and
      Song, Xingyi",
    editor = "Calzolari, Nicoletta  and
      Kan, Min-Yen  and
      Hoste, Veronique  and
      Lenci, Alessandro  and
      Sakti, Sakriani  and
      Xue, Nianwen",
    booktitle = "Proceedings of the 2024 Joint International Conference on Computational Linguistics, Language Resources and Evaluation (LREC-COLING 2024)",
    month = may,
    year = "2024",
    address = "Torino, Italia",
    publisher = "ELRA and ICCL",
    url = "https://aclanthology.org/2024.lrec-main.887/",
    pages = "10160--10171"
}

@inproceedings{papineni2002bleu,
  title={Bleu: a method for automatic evaluation of machine translation},
  author={Papineni, Kishore and Roukos, Salim and Ward, Todd and Zhu, Wei-Jing},
  booktitle={Proceedings of the 40th annual meeting of the Association for Computational Linguistics},
  pages={311--318},
  year={2002}
}

@misc{alvisi2025mappingitaliantelegramecosystem,
      title={Mapping the Italian Telegram Ecosystem: Communities, Toxicity, and Hate Speech},
      author={Lorenzo Alvisi and Serena Tardelli and Maurizio Tesconi},
      year={2025},
      eprint={2504.19594},
      archivePrefix={arXiv},
      primaryClass={cs.SI},
      url={https://arxiv.org/abs/2504.19594},
}

@inproceedings{xiaobaaiembeddings,
author = {Xiao, Shitao and Liu, Zheng and Zhang, Peitian and Muennighoff, Niklas and Lian, Defu and Nie, Jian-Yun},
title = {C-Pack: Packed Resources For General Chinese Embeddings},
year = {2024},
isbn = {9798400704314},
publisher = {Association for Computing Machinery},
address = {New York, NY, USA},
url = {https://doi.org/10.1145/3626772.3657878},
doi = {10.1145/3626772.3657878},
booktitle = {Proceedings of the 47th International ACM SIGIR Conference on Research and Development in Information Retrieval},
pages = {641–649},
numpages = {9},
location = {Washington DC, USA},
series = {SIGIR '24}
}

@article{lewis2020retrieval,
  title={Retrieval-augmented generation for knowledge-intensive nlp tasks},
  author={Lewis, Patrick and Perez, Ethan and Piktus, Aleksandra and Petroni, Fabio and Karpukhin, Vladimir and Goyal, Naman and K{\"u}ttler, Heinrich and Lewis, Mike and Yih, Wen-tau and Rockt{\"a}schel, Tim and others},
  journal={Advances in neural information processing systems},
  volume={33},
  pages={9459--9474},
  year={2020}
}

@misc{datafiniti_hotel_reviews_2016,
  author       = {{Datafiniti}},
  title        = {Hotel Reviews Dataset},
  howpublished = {Kaggle dataset},
  year         = {2016},
  note         = {Available at \url{https://www.kaggle.com/datasets/datafiniti/hotel-reviews}},
}

@misc{qwen2025qwen25technicalreport,
      title={Qwen2.5 Technical Report}, 
      author={Qwen},
      year={2025},
      eprint={2412.15115},
      archivePrefix={arXiv},
      primaryClass={cs.CL},
      url={https://arxiv.org/abs/2412.15115}, 
}

@misc{openai2024gpt4ocard,
  title        = {GPT-4o System Card},
  author       = {{OpenAI}},
  year         = {2024},
  eprint       = {2410.21276},
  archivePrefix= {arXiv},
  primaryClass = {cs.CL},
  url          = {https://arxiv.org/abs/2410.21276}
}

@misc{businessofapps_telegram_2025,
  author       = {{Business of Apps}},
  title        = {Telegram Statistics},
  howpublished = {\url{https://www.businessofapps.com/data/telegram-statistics/}},
  note         = {Accessed: 2025-01-27}
}

@misc{telegram_tos_2025,
  author  = {{Telegram}},
  title   = {Terms of Service},
  year    = {2025},
  url     = {https://telegram.org/tos/},
  note    = {Accessed: 2025-01-27}
}

@misc{telegram_api_terms_2025,
  author = {{Telegram}},
  title  = {Telegram API Terms of Service},
  year   = {2025},
  url    = {https://core.telegram.org/api/terms},
  note   = {Accessed: 2025-01-27}
}

@misc{gdpr_article85_2025,
  author = {{GDPR-Text.com}},
  title  = {Article 85 GDPR},
  year   = {2025},
  url    = {https://gdpr-text.com/read/article-85/},
  note   = {Accessed: 2025-01-27}
}

@misc{mistral-nemo,
  author = {{Mistral AI}},
  title  = {Mistral Nemo},
  year   = {2025},
  url    = {https://mistral.ai/news/mistral-nemo},
  note   = {Accessed: 2025-01-27}
}

@misc{perspective_api,
  author = {{Perspective API}},
  title  = {Perspective API},
  year   = {2025},
  url    = {https://www.perspectiveapi.com/},
  note   = {Accessed: 2025-01-27}
}
\bibliographystyle{acl_natbib}
\clearpage
\appendix
\section{Prompts used}
\subsection{Prompts used for generating three JSONs for each file}\label{app:a1}
The text placeholder may contain either Telegram messages or hotel reviews. In place of the example placeholder, three realistic examples illustrating potential Telegram QA or Hotel Review QA outputs are provided, though they are omitted here for readability.

\begin{tcolorbox}[
    enhanced,
    colback=gray!5,        % sfondo del contenuto
    colframe=black,        % colore del bordo esterno
    title=Telegram messages, % titolo con icona
    colbacktitle=gray!30,  % colore di sfondo del titolo
    coltitle=black,        % colore del testo nel titolo
    fonttitle=\bfseries,   % titolo in grassetto
    boxed title style={sharp corners}, % stile angoli del titolo
    attach boxed title to top left={yshift=-2mm, xshift=5mm}, % posizione del titolo
    boxrule=0.5pt          % spessore del bordo
]
You have a series of social media messages that you want to use to generate a relevant question and a simple, verifiable answer.\\
Each message is separated by "*************** Message \textbf{\{message\_id\}} ***************". \\
Please analyze these messages and create:\\
- A clear and relevant question based on the information in only one of these messages, avoiding any reference to the messages or the message number.\\
- A concise and simple answer (preferably a few words) that can be easily verified, avoiding unnecessary complexity.\\
- Ensure that the answer to the question is unique within all the messages. \\
There should be no other message that could provide a similar or alternative answer to the question. \\
The answer must come only from the specific message indicated.\\ \\

As an example: "if in the text OF ONE MESSAGE is mentioned that mountains are red ( if red is the only possible color mentioned for mountains in ALL messages), then a good pair question/answer will be: " Q: what is the color of mountains? A: red;"\\ \\

The output should be only the JSON format, containing at least 3 examples of questions and answers and the relevant message number in which the answer is present.

\textbf{\{text\}}

Below is an example of the JSON output:
\textbf{\{examples\}}
\end{tcolorbox}

\begin{tcolorbox}[
    enhanced,
    colback=gray!5,        % sfondo del contenuto
    colframe=black,        % colore del bordo esterno
    title=Hotel Reviews, % titolo con icona
    colbacktitle=gray!30,  % colore di sfondo del titolo
    coltitle=black,        % colore del testo nel titolo
    fonttitle=\bfseries,   % titolo in grassetto
    boxed title style={sharp corners}, % stile angoli del titolo
    attach boxed title to top left={yshift=-2mm, xshift=5mm}, % posizione del titolo
    boxrule=0.5pt          % spessore del bordo
]
You have a series of hotel reviews that you want to use to generate a relevant question and a simple, verifiable answer.\\
Each review is separated by "********** Review \textbf{\{review\_id\}} **********".\\
Please analyze these reviews and create:\\
- A clear and relevant question based on the information in only one of these reviews, avoiding any reference to the reviews, the review number or the hotel title.\\
- A concise and simple answer (preferably a few words) that can be easily verified, avoiding unnecessary complexity.\\
- Ensure that the answer to the question is unique within all the reviews. \\
There should be no other review that could provide a similar or alternative answer to the question. \\
The answer must come only from the specific review indicated.\\ \\

As an example: "if in the text OF ONE REVIEW is mentioned that mountains are red ( if red is the only possible color mentioned for mountains in ALL messages), then a good pair question/answer will be: " Q: what is the color of mountains? A: red;"\\ \\

The output should be only the JSON format, containing at least 3 examples of questions and answers and the relevant review number in which the answer is present.

\textbf{\{text\}}

Below is an example of the JSON output:
\textbf{\{examples\}}
\end{tcolorbox}

\subsection{Prompt used for select single QA between the three pairs generated}\label{app:a2}
The \textbf{type} placeholder can be either a message(s) or review(s) string, depending on the QA being created. The \textbf{text} placeholder contains the three JSON objects generated in the previous step, while the \textbf{example} placeholder contains an output example, which is omitted here for brevity.
\begin{tcolorbox}[
    enhanced,
    colback=gray!5,        % sfondo del contenuto
    colframe=black,        % colore del bordo esterno
    title=, % titolo con icona
    colbacktitle=gray!30,  % colore di sfondo del titolo
    coltitle=black,        % colore del testo nel titolo
    fonttitle=\bfseries,   % titolo in grassetto
    boxed title style={sharp corners}, % stile angoli del titolo
    attach boxed title to top left={yshift=-2mm, xshift=5mm}, % posizione del titolo
    boxrule=0.5pt          % spessore del bordo
]
Given a list of \textbf{\{type\}} and a JSON containing 3 question/answer pairs with the \textbf{\{type\}} from which the answer is derived, select the JSON object that contains the shortest and simplest answer, ensuring that it is correct and is the only possible answer based on the information contained in all the other \textbf{\{type\}}.\\

\textbf{\{text\}}\\

Returns only the selected json.\\
Below is an example of the JSON output:\\
\textbf{\{example\}}
\end{tcolorbox}

\subsection{Prompts used for the experiment on the AMAQA benchmark}
Here we report the prompts used by the LLMs to answer the questions of the AMAQA benchmark.

\begin{tcolorbox}[
    enhanced,
    colback=gray!5,        % sfondo del contenuto
    colframe=black,        % colore del bordo esterno
    title=System prompt, % titolo con icona
    colbacktitle=gray!30,  % colore di sfondo del titolo
    coltitle=black,        % colore del testo nel titolo
    fonttitle=\bfseries,   % titolo in grassetto
    boxed title style={sharp corners}, % stile angoli del titolo
    attach boxed title to top left={yshift=-2mm, xshift=5mm}, % posizione del titolo
    boxrule=0.5pt          % spessore del bordo
]
You are an assistant for question-answering tasks.\\
\hspace*{0.2cm}Instructions: \\
\hspace*{0.4cm}1. Familiarize yourself with the provided context.\\
\hspace*{0.4cm}2. If a document doesn't seem relevant with respect to the question, omit it from your response. \\
\hspace*{0.4cm}3. Never refer to facts that are not in your context. \\
\hspace*{0.4cm}4. Your answers should be relevant.\\
\hspace*{0.4cm}5. If your answer is a date and a time, please write it in the format "YYYY-MM-DD HH:MM:SS".\\
\hspace*{0.4cm}6. If you are asked for specific information (for example, a name, a place, or a date) answer directly with that information, without adding any other information.\\ \\

\hspace*{0.4cm}Note:\\
\hspace*{0.6cm}You cannot refer to facts or events if they are not in your context.
\end{tcolorbox}

\begin{tcolorbox}[
    enhanced,
    colback=gray!5,        % sfondo del contenuto
    colframe=black,        % colore del bordo esterno
    title=User prompt, % titolo con icona
    colbacktitle=gray!30,  % colore di sfondo del titolo
    coltitle=black,        % colore del testo nel titolo
    fonttitle=\bfseries,   % titolo in grassetto
    boxed title style={sharp corners}, % stile angoli del titolo
    attach boxed title to top left={yshift=-2mm, xshift=5mm}, % posizione del titolo
    boxrule=0.5pt          % spessore del bordo
]
Context: \textbf{\{context\}}\\
Question: \textbf{\{question\}}\\ \\
Please analyze the provided context and answer the question using a concise and simple answer (preferably a few words or a very short phrase) that can be easily verified, avoiding unnecessary complexity.\\
If you are asked for specific information (for example, a name, a place, or a date) answer directly with that information, without adding any other information.\\
If your answer is a date and a time, please write it in the format "YYYY-MM-DD HH:MM:SS".\\
The answer must come only from the context.\\ \\
Answer:
\end{tcolorbox}

\section{Annotators guidelines}\label{app:criteria}
Annotators should follow these guidelines to discard or
validate Question-Answer pairs:
\begin{itemize}
    \item \textit{Correctness of the question and answer:} The question had to be clear and meaningful, while the answer needed to be accurate and consistent with the document from which it was extracted.  
    \item \textit{Clarity and completeness:} The QA pair was required to avoid ambiguity (i.e., there should not be multiple plausible answers to the same question) and the answer needed to be complete, without omitting relevant information.  
    \item \textit{Originality with respect to LLMs:} To ensure the QA pairs were novel and not already known to LLMs, the question (or a minimally altered version of it) was submitted to ChatGPT. QA pairs were excluded if the model was able to provide the correct answer without relying on supporting documents.  
\end{itemize}
\clearpage
\section{Complementary Results}\label{app:c}
This section reports additional tables and figures presenting complementary results that support the analyses discussed in the main text. These materials were not included in the primary manuscript for readability and conciseness reasons, but are provided here to ensure completeness of the experimental evaluation.

\begin{figure}[ht]
  \centering
  \includegraphics[width=0.48\textwidth]{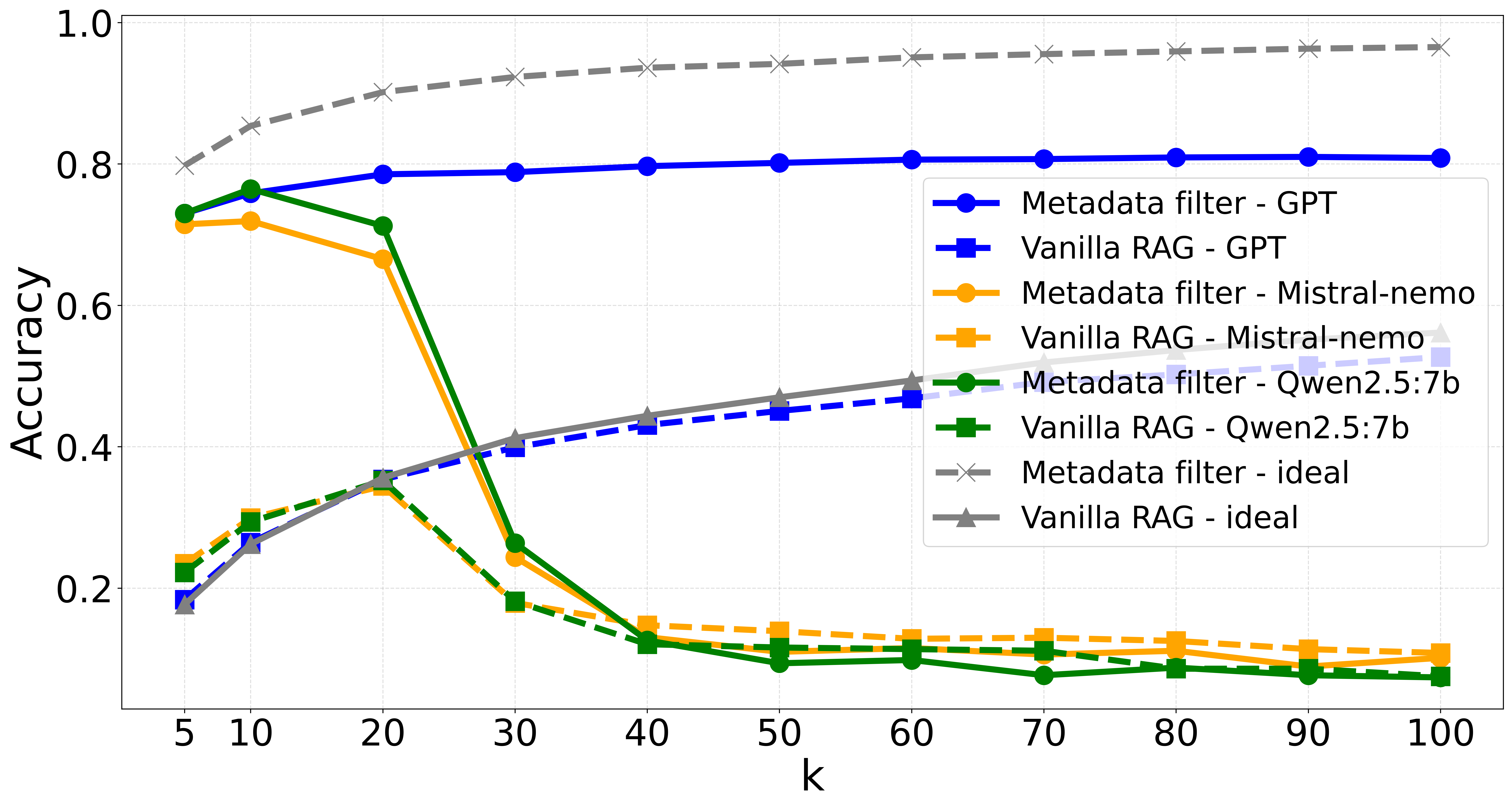}
  \caption{RAG Accuracy  comparison between Vanilla RAG and Metadata filter on Telegram subset} 
  \label{fig:accuracy_tg}
\end{figure}

\begin{figure}[ht]
  \centering
      \includegraphics[width=0.48\textwidth]{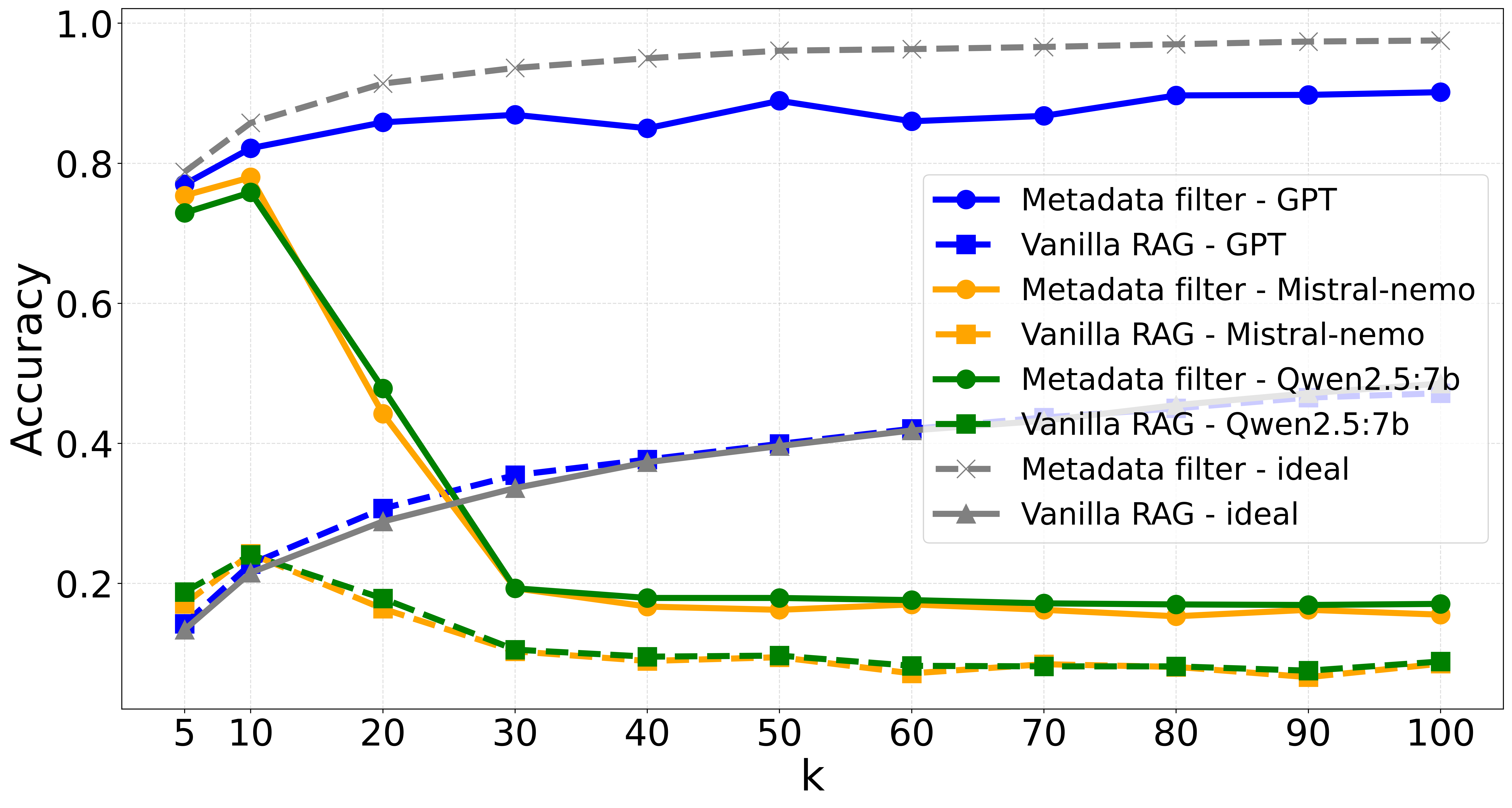}
  \caption{RAG Accuracy  comparison between Vanilla RAG and Metadata filter on Hotel Reviews subset} 
  \label{fig:accuracy_hr}
\end{figure}

\begin{figure}[ht]
  \centering
  \includegraphics[width=0.48\textwidth]{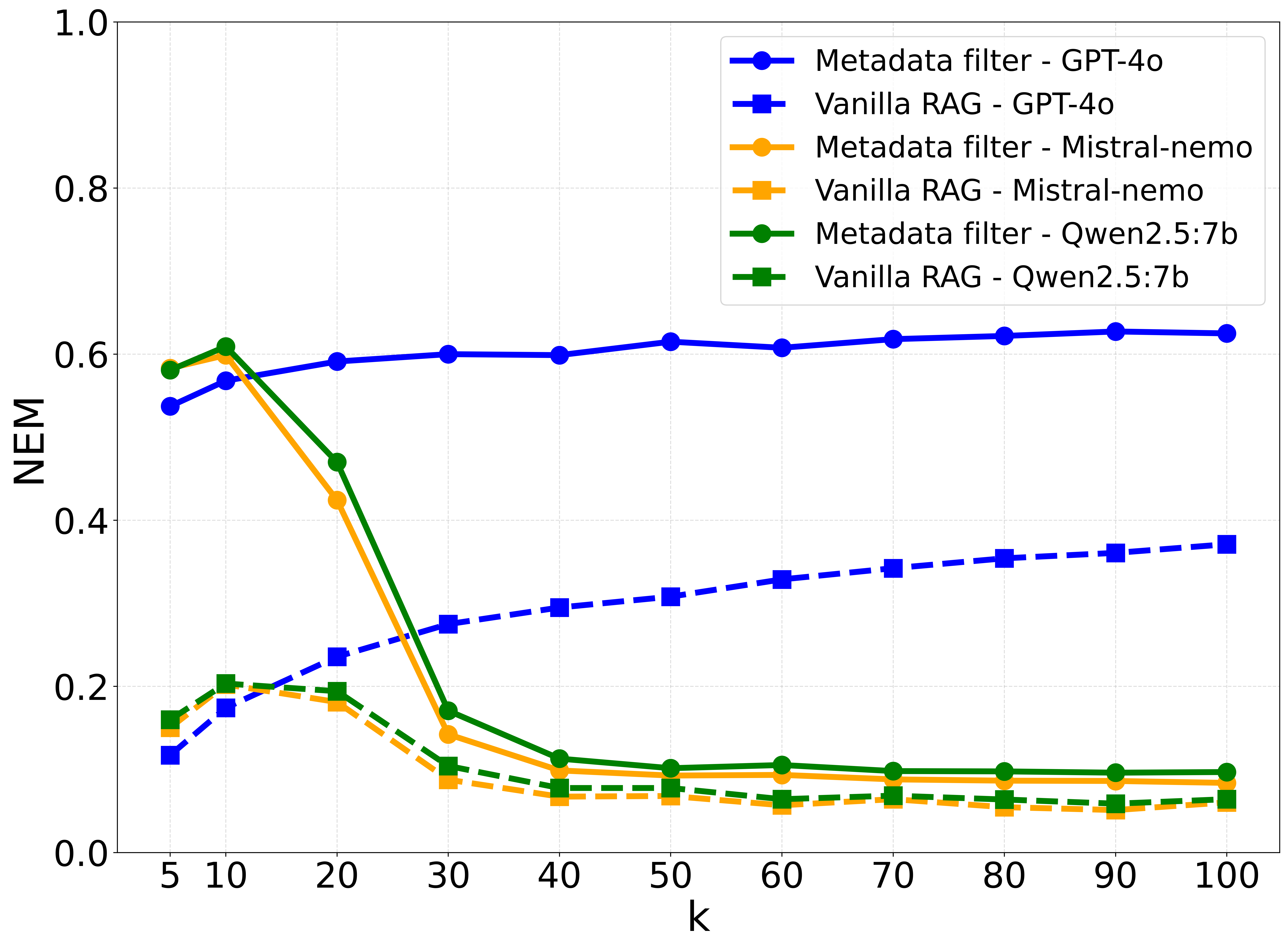}
  \caption{Comparison of RAG Normalized Exact Match between Vanilla RAG and RAG with metadata filter.} 
  \label{fig:nem_2}
\end{figure}

\begin{figure}[ht]
  \centering
  \includegraphics[width=0.48\textwidth]{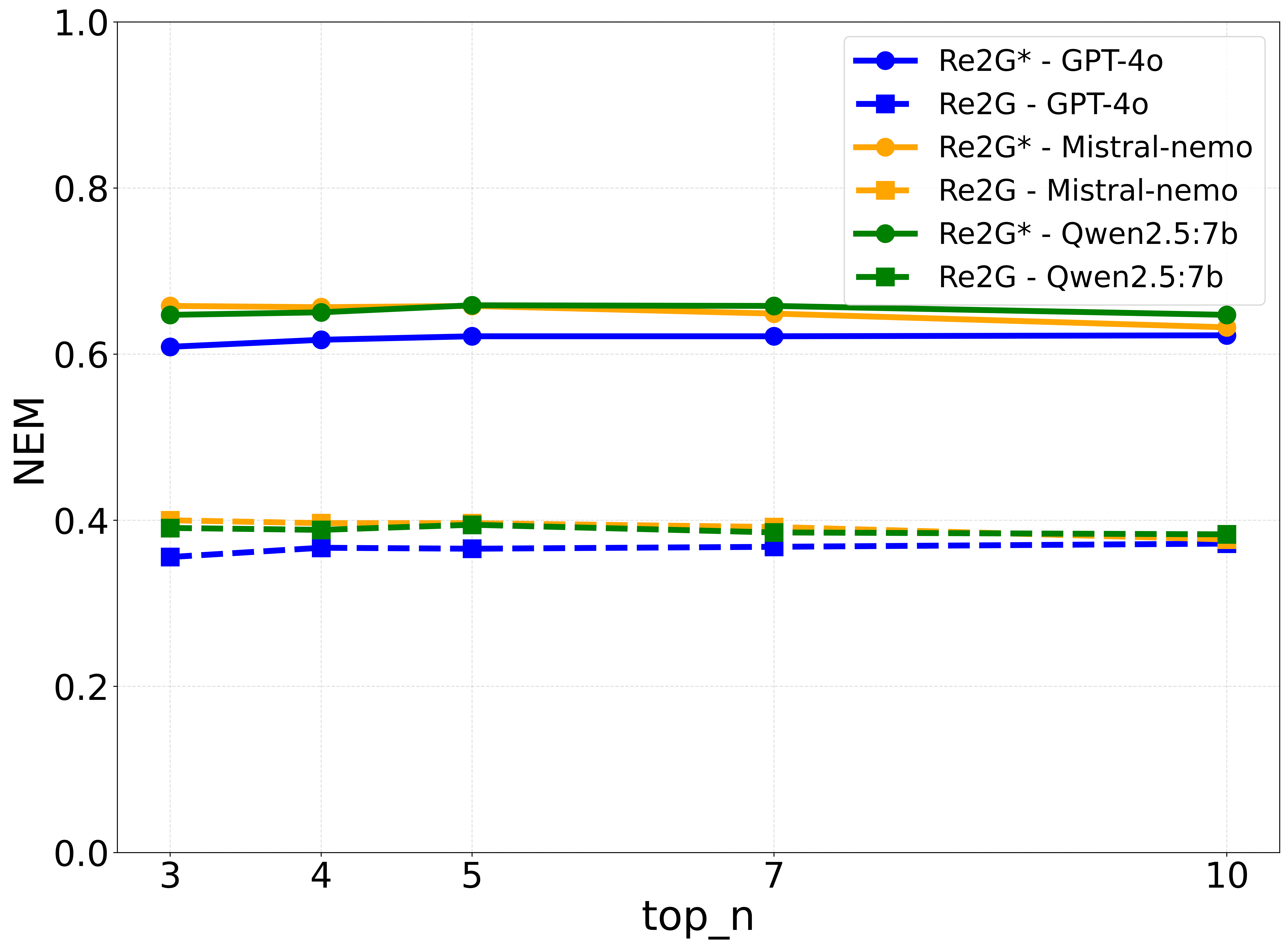}
  \caption{Comparison of Normalized Exact Match between RAG architectures using rerankers.} 
  \label{fig:nem_1}
\end{figure}

\begin{table*}[ht]
\begin{center}
\begin{tabular}{l|c|c|c|c|c|c}
\toprule
\textit{Subset} & \textit{RAG architecture} & \textit{LLM} & \textit{Acc.} & \textit{NEM} & \textit{k} & \textit{top\_n} \\
\midrule
Telegram & $\text{Re}^2\text{G}$* & GPT-4o & 0.81 & 0.57  & 100 & 10 \\
Telegram & Metadata filtering & GPT-4o & 0.81 & 0.57  & 100 & - \\
Telegram &$\text{Re}^2\text{G}$ & GPT-4o & 0.53 & 0.36 & 100 & 10 \\
Telegram & Vanilla RAG & GPT-4o & 0.53 & 0.38 & 100 & - \\
\midrule
Telegram & $\text{Re}^2\text{G}$* & Qwen2.5:7b & 0.80 & 0.66  & 100 & 7 \\
Telegram &Metadata filtering & Qwen2.5:7b & 0.76 & 0.62  & 10 & - \\
Telegram &$\text{Re}^2\text{G}$ & Qwen2.5:7b & 0.54 & 0.41 & 100 & 5 \\
Telegram & Vanilla RAG & Qwen2.5:7b & 0.35 & 0.26 & 20 & - \\
\midrule
Hotel Reviews & $\text{Re}^2\text{G}$* & GPT-4o & 0.89 & 0.68  & 100 & 10 \\
Hotel Reviews & Metadata filtering & GPT-4o & 0.90 & 0.68  & 100 & - \\
Hotel Reviews &$\text{Re}^2\text{G}$ & GPT-4o & 0.48 & 0.38 & 100 & 10 \\
Hotel Reviews & Vanilla RAG & GPT-4o & 0.47 & 0.36 & 100 & - \\
\midrule
Hotel Reviews & $\text{Re}^2\text{G}$* & Mistral-nemo & 0.87 & 0.66  & 100 & 3 \\
Hotel Reviews &Metadata filtering & Mistral-nemo & 0.78 & 0.61  & 10 & - \\
Hotel Reviews &$\text{Re}^2\text{G}$ & Mistral-nemo & 0.48 & 0.37 & 100 & 5 \\
Hotel Reviews & Vanilla RAG & Mistral-nemo & 0.24 & 0.18 & 10 & - \\
\bottomrule
\end{tabular}
\caption{Comparison of RAG architectures under optimal settings based on accuracy (acc.) and NEM.}
\label{tab:total_res}
\end{center}
\end{table*}

\end{document}